\documentclass[letterpaper, 10 pt, journal]{IEEEtran}
\IEEEoverridecommandlockouts
\usepackage[english]{babel}

\usepackage{amsmath, amssymb, amsthm}
\usepackage{cite}
\usepackage{xcolor}
\usepackage{stmaryrd}
\usepackage{threeparttable}
\usepackage{booktabs}
\usepackage[caption=false]{subfig}

\usepackage[linesnumbered,ruled,vlined]{algorithm2e}
\usepackage{optidef}

\usepackage{enumitem}
\usepackage{url}
\usepackage{accents}
\usepackage[version=4]{mhchem}
\usepackage{soul} 

\newtheorem{remark}{Remark}

\def\matt#1{\begin{bmatrix}#1\end{bmatrix}}

\def\BibTeX{{\rm B\kern-.05em{\sc i\kern-.025em b}\kern-.08em
    T\kern-.1667em\lower.7ex\hbox{E}\kern-.125emX}}

\usepackage{tabularx}

\usepackage[hidelinks]{hyperref}
\hypersetup{
    colorlinks=true,
    linkcolor=blue,
    filecolor=magenta,
    urlcolor=cyan,
    citecolor=red
}

\begin{document}

\title{
Integrated Optimal Fast Charging and Active Thermal Management of Lithium-Ion Batteries in Extreme Ambient Temperatures
}

\author{Zehui Lu, Hao Tu, Huazhen Fang, Yebin Wang, Shaoshuai Mou
\thanks{Z. Lu and S. Mou are with the School of Aeronautics and Astronautics, Purdue University, West Lafayette, IN 47907, USA {\tt\small \{lu846,mous\}@purdue.edu} }
\thanks{H. Tu and H. Fang are with the Department of Mechanical Engineering, University of Kansas, Lawrence, KS 66045, USA {\tt\small \{tuhao,fang\}@ku.edu} }
\thanks{Y. Wang is with Mitsubishi Electric Research Laboratories, Cambridge, MA 02139, USA {\tt\small yebinwang@ieee.org} }
}

\maketitle

\begin{abstract}
This paper presents an integrated control strategy for optimal fast charging and active thermal management of Lithium-ion batteries in extreme ambient temperatures, striking a balance between charging speed and battery health.
A control-oriented thermal-NDC (nonlinear double-capacitor) battery model is proposed to describe the electrical and thermal dynamics, incorporating the effects of both an active thermal source and ambient temperature.
A state-feedback model predictive control algorithm is then developed for optimal fast charging and active thermal management. Numerical experiments validate the algorithm under extreme temperatures, showing that the proposed algorithm can energy-efficiently adjust the battery temperature, thereby balancing charging speed and battery health.
Additionally, an output-feedback model predictive control algorithm with an extended Kalman filter is proposed for battery charging when states are partially measurable. Numerical experiments validate the effectiveness under extreme temperatures.
\end{abstract}


\section{Introduction} \label{sec:intro}




Lithium-ion batteries (LiBs) have become widely accepted in various fields, including electrified transportation \cite{khaligh2010battery}, consumer electronics \cite{goodenough2013li}, and renewable energy \cite{goodenough2015energy}, due to their favorable characteristics such as high voltage and power density, low self-discharge rates, and lack of memory effects \cite{wang2017revisiting}.
However, a critical concern for LiBs is their lifespan, which is highly sensitive to battery deterioration. Factors influencing longevity include charging techniques \cite{ramadass2004development} and operating temperature \cite{wang2016lithium,yang2018fast}. Research has increasingly focused on advanced battery energy management to enhance LiB performance, safety, and durability.

The suitability of LiBs for specific applications is limited by concerns about battery charging techniques and operating temperature. For instance, electric vertical takeoff and landing (eVTOL) aircraft typically require batteries with a discharging rate three times that of electric vehicles (EVs) during takeoff and landing, as well as four times the fast charging frequency of EVs \cite{yang2021challenges}.
Charging and discharging LiBs in such systems can generate significant heat, which can pose a risk to the entire system, not to mention the existing frequent fires during EV charging \cite{sun2020review}.
Moreover, as electrified systems have been deployed in wildly diversified areas, managing battery charging quickly and safely, even under extreme ambient temperatures, has become an imperative research topic. Therefore, this paper aims to investigate an integrated strategy for LiB optimal fast charging and thermal management under extreme temperatures, striking a balance between charging speed and battery health.

\subsection{Literature Review}

Optimal LiB charging relies on accurate and comprehensive modeling of battery dynamics, including electrical, thermal, and aging aspects.
Two primary categories of LiB models are widely recognized.
1) \emph{Electrochemical Models}: These models are derived from electrochemical principles and aim to explain the electrochemical reactions and physical phenomena occurring inside a battery cell during charging and discharging. They are typically characterized by high-order partial differential equations \cite{klein2012electrochemical,weaver2020novel}.
2) \emph{Equivalent Circuit Models (ECMs)}: In contrast, ECMs replicate a battery's current-voltage characteristics by utilizing electrical circuits consisting of resistors, capacitors, and voltage sources \cite{tian2020nonlinear}, offering excellent computational efficiency and making them particularly suitable for real-time battery energy management.

A fundamental ECM, commonly known as the Rint (internal resistance) model, consists of an open-circuit voltage (OCV) source cascaded with an internal resistor, where the voltage source is state-of-charge (SoC)-dependent \cite{plett2015battery}.
To describe the transient voltage response within a cell, the Rint model can be expanded by adding some serially connected resistor–capacitor (RC) pairs, leading to the Thevenin model \cite{plett2015battery}.
When multiple RC circuits are integrated into the Thevenin model, it evolves into the Dual Polarization (DP) model, which captures multi-timescale voltage transients during charging and discharging \cite{he2011evaluation}.
Another ECM gaining attention is the double-capacitor model \cite{johnson2002battery}, comprising two capacitors in parallel, representing the bulk inner portion and surface region of an electrode, respectively.
This model describes charge diffusion and storage mechanisms in a battery's electrode \cite{fang2016health}.
Unlike the Thevenin and DP models, this circuit structure accounts for rate capacity effects and charge recovery phenomena, making it appealing for battery charging control \cite{fang2016health}.
However, this model, being linear, struggles to capture nonlinear battery phenomena such as the nonlinear SoC-OCV relation.
To address these limitations, a Nonlinear Double-Capacitor (NDC) model is proposed in \cite{tian2020nonlinear}, effectively capturing the battery's nonlinear behaviors by introducing a nonlinear-mapping-based voltage source and a series RC circuit.

Regarding the thermal dynamics of LiB, Lin et al. \cite{lin2014lumped} developed an electro-thermal model for cylindrical batteries, consisting of two sub-models: a DP model and a two-state thermal model. The thermal model represents the dynamics of the battery surface and core temperatures, while the DP model characterizes electrochemical processes. These models are interconnected through heat generation and temperature-dependent electric parameters.
Perez et al. \cite{perez2017optimal} expanded on the work \cite{lin2014lumped} by creating an electro-thermal-aging battery model. In this model, the electric and aging sub-models are influenced by the battery core temperature, as captured by the two-state thermal sub-model.
Biju and Fang \cite{biju2023battx} proposed an electro-thermal model that combines an ECM with a single particle model for electrolyte and thermal dynamics. This model uses multiple circuits to simulate a LiB cell's electrode, electrolyte, and thermal dynamics, considering their effects on terminal voltage. Consequently, the model accurately approximates major electrochemical and physical processes during charging and discharging.


Over the past two decades, there has been continuous attention on exploring suitable methods for charging LiBs.
One of the most common approaches in the industry is constant-current/constant-voltage (CC/CV) charging.
This method involves applying a constant current to charge LiB cells until they reach a specific voltage threshold \cite{hussein2011review}.
Subsequently, a constant voltage is maintained to charge the cell with a gradually decreasing current \cite{hussein2011review}.
However, these model-free methods typically rely on tuning some heuristic-based charging parameters, offering empirical or conservative assurances for charging safety and speed.

Researchers have been developing model-based charging strategies by integrating physics-based LiB models with optimization techniques to achieve faster charging.
Nonlinear model predictive control (MPC) has gained significant attention for this purpose, as it can handle nonlinear objectives, system dynamics, and constraints related to the state and control of the entire system \cite{10155935}.
However, solving a nonlinear MPC problem at run-time can be computationally expensive, especially with a longer prediction horizon.
To reduce the computational load, Klein et al. \cite{klein2011optimal} formulate a nonlinear MPC charging problem with a 1-D electrochemical model of LiBs and only one-step prediction.
Another approach to reducing the computational load is model reduction, often used in literature to simplify a battery model and enable real-time MPC strategies.
Ref. \cite{xavier2015lithium,zou2017electrothermal} formulate a linear MPC problem based on the linear Thevenin model.
Fang et al. \cite{fang2016health} employ linear quadratic control to achieve health-aware charging using the linear double-capacitor model.
Zou et al. \cite{zou2018model} linearize a nonlinear electrochemical model along a reference SoC trajectory and then solve an optimal tracking problem at run-time.
A hierarchical MPC strategy in \cite{ouyang2018optimal} generates a reference current trajectory at a slow time scale and performs current reference tracking at a faster time scale, reducing the run-time computational load.

With the increase in computational power over the past decade, more literature has emerged on charging strategies to achieve specific objectives related to LiB charging speed, safety, health, aging, etc.
Perez et al. \cite{perez2017optimal} propose a multi-objective optimal charging control based on the electro-thermal-aging battery model mentioned earlier. The objective function is a linear combination of total charging time and the loss of the battery's state-of-health (SOH).
Based on an RC-based linear equivalent circuit model for a LiB cell, Fang et al. \cite{fang2018optimal} optimize both the magnitudes and duty cycles of current pulses to balance health considerations and charging rates.
Tian et al. \cite{tian2020real} utilize explicit MPC to achieve real-time optimal charging control. They propose a health-aware constraint on the voltages on two capacitors to limit the battery's internal stress during charging.
Azimi et al. \cite{azimi2022extending} formulate and solve a multi-objective optimal control problem for a LiB module made of series-connected cells, aiming for fast charging while minimizing degradation.

In addition, recent literature has been exploring several emerging techniques for real-time battery charging control.
For example, a two-level architecture is proposed in \cite{romagnoli2019feedback, goldar2020low, couto2021faster}. This architecture employs a low-level linear quadratic regulator to ensure stability and fast-tracking of a reference, while a high-level reference governor \cite{garone2017reference} enforces constraint satisfaction by adjusting the reference.
Feng et al. \cite{feng2024safe} apply cascade-CBF (control barrier function) to modify the charging control from the CC/CV strategy, using the same model as in \cite{lin2014lumped}.
This approach prevents violations of voltage, SoC, and battery surface temperature constraints.
Robust MPC is utilized in \cite{galuppini2024efficient,dong2024optimal} to design optimal control strategies that prevent constraint violations under system and process uncertainties.
Li et al. \cite{li2023nonlinear} converts a constrained optimal control problem into an output tracking problem through nonlinear model inversion-based control, thereby improving real-time computation performance.

Among the various aspects of LiB characteristics, safety and health (or aging) during fast charging are considered the most important factors.
Many studies, e.g. \cite{fang2016health,perez2017optimal,tian2020real,azimi2022extending,romagnoli2019feedback, goldar2020low, couto2021faster,feng2024safe,galuppini2024efficient,dong2024optimal,li2023nonlinear}, investigate how to incorporate safety and health-related constraints into optimization frameworks. These constraints ensure that the state and input at each time instance do not violate safety and health constraints.
Proper constraint formulation is crucial for developing model-based charging strategies to maintain battery safety and health.


To summarize the literature review, there is a gap in modeling LiB systems, particularly in capturing the nonlinear electro-thermal dynamics where electrical dynamics are influenced by thermal dynamics, and vice versa.
Existing thermal models also lack a description of how external heat sources affect the battery beyond ambient heat convection.
This paper aims to address these gaps by proposing a model that incorporates nonlinear electro-thermal dynamics and accounts for external heat sources.
The modeling of external heat sources could potentially enable battery charging under extreme ambient temperatures.
Additionally, the paper aims to investigate a model-based charging strategy to ensure safe, health, and fast charging, as well as active battery thermal management, even under extreme ambient temperatures.




\subsection{Contributions, Organization, and Notations}

This paper presents an integrated control strategy for optimal fast charging and active thermal management of LiBs in extreme ambient temperatures.
A control-oriented thermal-NDC battery model is proposed to describe the electrical and thermal dynamics, accounting for the impact from both an active thermal source and ambient temperature.
The thermal-NDC model enables the development of a state-feedback MPC algorithm, which integrates optimal fast charging and active thermal management for LiBs under extreme ambient temperatures.

Numerical experiments validate that the proposed algorithm under extreme temperatures can energy-efficiently adjust the battery temperature, thereby balancing charging speed and battery health.
Several insights are revealed and summarized in Section~\ref{sec:num_sim} to explain why the decisions made by the proposed algorithm lead to energy-efficient and health-aware optimal fast charging. The observations and explanations are consistent with the literature.
Additionally, an output-feedback MPC algorithm with an extended Kalman filter (EKF) is proposed for battery charging when states are partially measurable. Numerical experiments validate the effectiveness under extreme temperatures.
The contributions of this paper are summarized as follows:
\begin{enumerate}
\item a thermal-NDC model for control-oriented battery dynamics;
\item a state-feedback MPC algorithm;
\item an output-feedback MPC algorithm with battery state estimation from an EKF.
\end{enumerate}

The rest of this paper is organized as follows:
\begin{enumerate}
\item Section~\ref{sec:system_model} proposes the thermal-NDC model;
\item Section~\ref{sec:problem} formulates a state-feedback MPC algorithm for battery charging and active thermal management;
\item Section~\ref{sec:num_sim} performs several numerical experiments to evaluate the performance of the proposed algorithm with different ambient temperatures and investigate some factors that affect charging performance;
\item Section~\ref{sec:ekf_mpc} proposes an EKF-based output-feedback MPC strategy and performs numerical experiments to evaluate its performance;
\item Section~\ref{sec:conclusion} concludes this paper and discusses limitations and future work.
\end{enumerate}

\emph{Notations. } The real number set is denoted by $\mathbb{R}$.
The natural number set is denoted by $\mathbb{N}$.
Let $\llbracket a,b \rrbracket$ denote a set of all integers between integers $a$ and $b$, with both ends included.
For $\boldsymbol{x}, \boldsymbol{y} \in \mathbb{R}^n$, $\boldsymbol{x} \leq \boldsymbol{y}$ indicates element-wise inequality.
Let $\text{col}\{ \boldsymbol{v}_1, \cdots, \boldsymbol{v}_a \}$ denote a column stack of elements $\boldsymbol{v}_1, \cdots, \boldsymbol{v}_a $, which may be scalars, vectors or matrices, i.e., $\text{col}\{ \boldsymbol{v}_1, \cdots, \boldsymbol{v}_a \} \triangleq {\matt{{\boldsymbol{v}_1}^{\top} & \cdots & {\boldsymbol{v}_a}^{\top}}}^{\top}$.
For a matrix $\boldsymbol{A} \in \mathbb{R}^{n \times m}$, $A[i,j]$ indicates the entry in the $i$-th row and the $j$-th column of $\boldsymbol{A}$, $i \in \llbracket 1, n \rrbracket$, $j \in \llbracket 1, m \rrbracket$.
Denote $\boldsymbol{I}_n$ as an identity matrix in $\mathbb{R}^{n \times n}$.
The exponential function is denoted by $\text{exp}(\cdot)$, i.e., $\text{exp}(x) \triangleq e^{x}$, where $x \in \mathbb{R}$.
$\text{diag}(a, \ b, \ \cdots, \ c)$ denote a square real matrix, where the non-diagonal elements are zeros and the diagonal elements are $a, b, \cdots, c \in \mathbb{R}$.

\section{Battery Electro-thermal Modeling} \label{sec:system_model}

This section presents a control-oriented battery model named the thermal-NDC (nonlinear double-capacitor) model, which contains a nonlinear NDC model \cite{tian2020nonlinear} and a two-state lumped thermal model, where the battery surface temperature is affected by the heat generated from the electro-thermal process within the battery, the heat diffusion between the ambient environment and the battery surface, and active thermal input.


\begin{figure}[ht]
\centering
\includegraphics[width=0.30\textwidth]{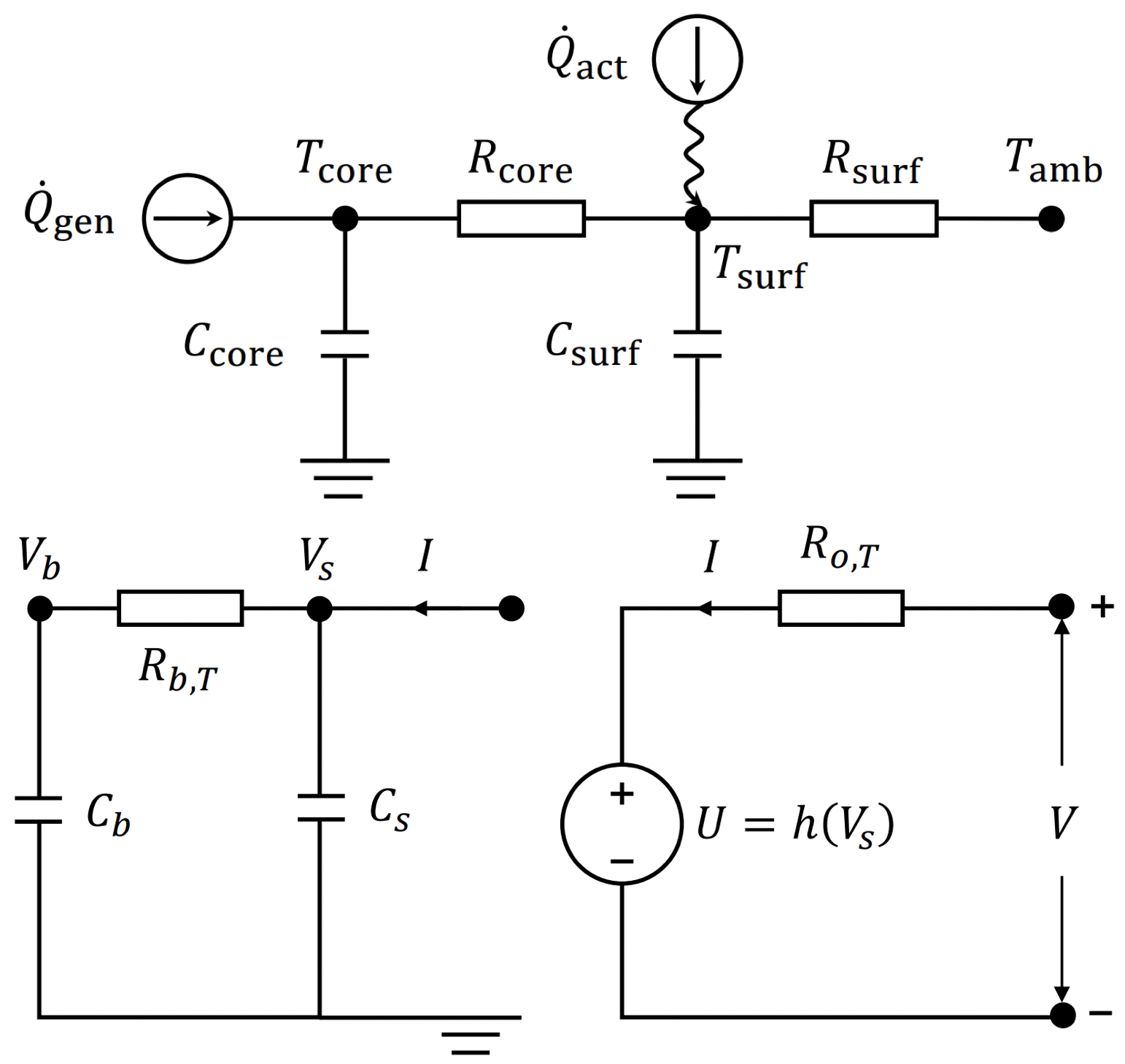}
\caption{The thermal-NDC (nonlinear double-capacitor) model. $V_{\mathrm{b}}$ and $V_{\mathrm{s}}$ are the normalized voltage in $[0\text{ V},1 \text{ V}]$.}
\label{fig:ndc_model}
\end{figure}

As shown in the lower portion of Fig. \ref{fig:ndc_model}, the thermal-NDC model uses electrical circuits to describe the diffusion and electrical process inside a Lithium-ion battery (LiB) cell. It contains two coupled sub-circuits. The first (left) circuit contains two parallel connected capacitors, $C_{\mathrm{b}}$ and $C_{\mathrm{s}}$, and one resistor $R_{\mathrm{b,T}}$. 
The migration of charge between $C_{\mathrm{b}}$ and $C_{\mathrm{s}}$ mimics the change of Lithium-ion concentrations within the electrode. 
Conceptually, $C_{\mathrm{b}}$ and $C_{\mathrm{s}}$ represent the electrode’s bulk inner region and surface region, respectively. 
The second (right) circuit contains two components in series, a voltage source $U$ and a resistor $R_{\mathrm{o,T}}$. Here, $U = h(V_{\mathrm{s}})$ serves as an open-circuit voltage source. $R_{\mathrm{o,T}}$ corresponds to the ohmic resistance and solid electrolyte interface resistance.
The first portion of the thermal-NDC model's governing equations is summarized in the following state-space form:
\begin{subequations} \label{eq:NDC_model}
\begin{align}
\matt{\dot{V}_{\mathrm{b}}(t) \\ \dot{V}_{\mathrm{s}}(t)} &= \boldsymbol{A}_1(t) \matt{V_{\mathrm{b}}(t) \\ V_{\mathrm{s}}(t)} + \boldsymbol{B}_1 I(t), \label{eq:NDC_model:two_v} \\
V(t) &= h(V_{\mathrm{s}}(t)) + R_{\mathrm{o, T}}(t) I(t), \label{eq:NDC_model:one_terminal_v}
\end{align}
\end{subequations}
where $V_{\mathrm{b}}(t)$ and $V_{\mathrm{s}}(t)$ are the normalized voltage in $[0\text{ V},1 \text{ V}]$ across $C_{\mathrm{b}}$ and $C_{\mathrm{s}}$, respectively; $I(t)$ is the input current; $I(t) < 0$ for discharging and $I(t) > 0$ for charging; $h: \mathbb{R} \mapsto \mathbb{R}$ denotes a mapping from $V_{\mathrm{s}}(t)$ to the open circuit voltage source, which is determined by experiments; the matrices $A_1(t)$ and $B_1$ are given by
\begin{equation} \label{eq:A1_B1}
\boldsymbol{A}_1(t) = \matt{ \dfrac{-1}{C_{\mathrm{b}} R_{\mathrm{b,T}}(t)} & \dfrac{1}{C_{\mathrm{b}} R_{\mathrm{b,T}}(t)} \\ \dfrac{1}{C_{\mathrm{s}} R_{\mathrm{b,T}}(t)} & \dfrac{-1}{C_{\mathrm{s}} R_{\mathrm{b, T}}(t)} }, \ \boldsymbol{B}_1 = \matt{0 \\ \dfrac{1}{C_{\mathrm{s}}}};
\end{equation}
$V(t)$ is the terminal voltage; \eqref{eq:NDC_model:one_terminal_v} indicates that the relation between the open-circuit voltage $U = h(V_{\mathrm{s}})$ and the terminal voltage $V$ is affected by the internal resistance $R_{\mathrm{o, T}}(t)$, which is dependent on both temperature and SoC.

The state of charge (SoC) is given by:
\begin{equation} \label{eq:soc_define}
\mathrm{SoC}(t) = \frac{ C_{\mathrm{b}} V_{\mathrm{b}}(t) + C_{\mathrm{s}} V_{\mathrm{s}}(t) }{ C_{\mathrm{b}} \overline{V}_{\mathrm{b}} + C_{\mathrm{s}} \overline{V}_{\mathrm{s}} } \times 100\%,
\end{equation}
where $\overline{V}_{\mathrm{b}}=1$ V and $\overline{V}_{\mathrm{s}} = 1$ V are the upper bounds of $V_{\mathrm{b}}(t)$ and $V_{\mathrm{s}}(t)$, respectively.
This paper aims to formulate a minimal state-space representation for the battery dynamics, thus SoC is not a state of the system but acts as another state coordinate of the states $V_{\mathrm{b}}$ and $V_{\mathrm{s}}$. In other words, according to \eqref{eq:soc_define}, one can determine SoC by $V_{\mathrm{b}}$ and $V_{\mathrm{s}}$, and replace SoC by \eqref{eq:soc_define}.
Here, $V_{\mathrm{b}}=V_{\mathrm{s}}=0$ V when the cell is depleted ($\mathrm{SoC}=0\%$), and $V_{\mathrm{b}}=V_{\mathrm{s}}=1$ V when the cell is fully charged ($\mathrm{SoC}=100\%$).

\begin{remark} \label{remark:normalize_v}
$V_{\mathrm{b}}$ and $V_{\mathrm{s}}$ must have an upper bound as a battery has a maximum capacity, and the upper bound could be arbitrary. $C_{\mathrm{b}}$ and $C_{\mathrm{s}}$ would vary accordingly so that $C_{\mathrm{b}} \overline{V}_{\mathrm{b}} + C_{\mathrm{s}} \overline{V}_{\mathrm{s}}$ is consistent with the battery's capacity. Without loss of generality and for simplicity of notation, $\overline{V}_{\mathrm{b}}=1$ \emph{V} and $\overline{V}_{\mathrm{s}} = 1$ \emph{V}.
\end{remark}

The internal resistance $R_{\mathrm{o,T}} \in \mathbb{R}$ is dependent on both temperature and SoC and is given by:
\begin{equation} \label{eq:internal_resis}
\begin{split}
&R_{\mathrm{o,T}}(t) = R_{\mathrm{o}}(\mathrm{SoC}(t)) \cdot \text{exp}(\kappa_1 (\frac{1}{T_{\mathrm{core}}(t)} - \frac{1}{T_{\mathrm{ref}}})), \\
&R_{\mathrm{o}}(\mathrm{SoC}(t)) = \gamma_1 + \gamma_2 \cdot \text{exp}(-\gamma_3 \mathrm{SoC}(t)),
\end{split}
\end{equation}
where $\kappa_1, \gamma_1, \gamma_2, \gamma_3 \in \mathbb{R}$ are battery-dependent parameters and determined by experiments;
$T_{\mathrm{core}} \in \mathbb{R}$ is the core temperature of the cell, which is later described in the lumped thermal model; $T_{\mathrm{ref}} \in \mathbb{R}$ is the reference temperature in the Arrhenius law.
The diffusion resistance $R_{\mathrm{b,T}} \in \mathbb{R}$ is also temperature-dependent and given by:
\begin{equation} \label{eq:diffusion_resis}
R_{\mathrm{b,T}}(t) = R_\mathrm{b} \cdot \text{exp}(\kappa_2 (\frac{1}{T_{\mathrm{core}}(t)} - \frac{1}{T_{\mathrm{ref}}})),
\end{equation}
where $\kappa_2 \in \mathbb{R}$ is a battery-dependent parameter and determined by experiments.

As shown in the upper portion of Fig. \ref{fig:ndc_model}, a two-state lumped thermal model, inspired by \cite{lin2014lumped}, is used to capture the thermal dynamics of a cylindrical LiB cell. The temperature along the cell's axial direction is assumed to be uniform. The cell's temperature distribution along the radial direction is simplified to two singular points which represent the core and surface.
The battery surface temperature is affected by $\dot{Q}_{\mathrm{act}}$ from active thermal input and the diffusion between the surface and the environment.
The second portion of the thermal-NDC model's governing equations is given by:
\begin{equation} \label{eq:lumped_thermal_model}
\matt{\dot{T}_{\mathrm{core}}(t) \\ \dot{T}_{\mathrm{surf}}(t)} = \boldsymbol{A}_2 \matt{T_{\mathrm{core}}(t) \\ T_{\mathrm{surf}}(t)} + \boldsymbol{B}_2 \matt{ \dot{Q}_{\mathrm{gen}}(t) \\ T_{\mathrm{amb}}(t) \\ \dot{Q}_{\mathrm{act}}(t) },
\end{equation}
where
\begin{equation*}
\begin{split}
\boldsymbol{A}_2 &= \matt{ \dfrac{-1}{R_{\mathrm{core}}C_{\mathrm{core}}} & \dfrac{1}{R_{\mathrm{core}}C_{\mathrm{core}}} \\ \dfrac{1}{R_{\mathrm{core}}C_{\mathrm{surf}}} & \dfrac{-1}{R_{\mathrm{surf}}C_{\mathrm{surf}}} + \dfrac{-1}{R_{\mathrm{core}}C_{\mathrm{surf}}} }, \\
\boldsymbol{B}_2 &= \matt{\dfrac{1}{C_{\mathrm{core}}} & 0 & 0 \\ 0 & \dfrac{1}{R_{\mathrm{surf}}C_{\mathrm{surf}}} & \dfrac{1}{C_{\mathrm{surf}}} }.
\end{split}
\end{equation*}
Here, $T_{\mathrm{core}}(t)$ and $T_{\mathrm{surf}}(t)$ are the temperature at the core and surface of the cell, respectively; $T_{\mathrm{amb}}(t)$ is the known ambient temperature at time $t$; $C_{\mathrm{core}}$ and $C_{\mathrm{surf}}$ are the cell's core and surface heat capacity. A thermal resistance $R_{\mathrm{core}}$ is used to model the conduction between the cell's core and surface. $R_{\mathrm{surf}}$ is the other thermal resistance that captures the convection between the cell's surface and the environment. In addition, the heat generation inside the cell is assumed to be concentrated in the core of the cell, which is considered to be
\begin{equation} \label{eq:heat_generation_rate}
\dot{Q}_{\mathrm{gen}}(t) = I(t) (V(t) - h(\mathrm{SoC}(t))).
\end{equation}
The effect of the heating and cooling systems is reflected in the heat exchange between the hot/cold plates to the cell's surface, i.e.
\begin{equation} \label{eq:heat_active}
\dot{Q}_{\mathrm{act}}(t) = \eta_{\mathrm{act}}P_{\mathrm{act}}(t),
\end{equation}
where $P_{\mathrm{act}}(t) > 0$ and $P_{\mathrm{act}}(t) < 0$ indicates the active heating and cooling power, respectively; 
$\eta_{\mathrm{act}} \in [0,1]$ is the efficiency coefficient.

\begin{remark}
For simplicity, \eqref{eq:heat_active} assumes the same efficiency for heating and cooling. Nevertheless, one can directly convert \eqref{eq:heat_active} into heating and cooling power with different efficiency. In detail, \eqref{eq:heat_active} can be written as
\begin{equation} \label{eq:heat_active_remark}
\dot{Q}_{\mathrm{act}}(t) = P_{\mathrm{neat}}(t) \equiv \eta_{\mathrm{heat}}P_{\mathrm{heat}}(t) + \eta_{\mathrm{cool}}P_{\mathrm{cool}}(t),
\end{equation}
where $P_{\mathrm{neat}}(t)$ is a decision variable. Let $P_{\mathrm{heat}}(t) \in [0, \overline{P}_{\mathrm{heat}}]$ and $P_{\mathrm{cool}}(t) \in [\underline{P}_{\mathrm{cool}}, 0]$. Then given the sign of $P_{\mathrm{neat}}(t)$, one can directly determine the value of $P_{\mathrm{heat}}(t)$ and $P_{\mathrm{cool}}(t)$. Given the power limit of $P_{\mathrm{heat}}(t)$ and $P_{\mathrm{cool}}(t)$, $P_{\mathrm{neat}}(t) \in [\eta_{\mathrm{cool}}\underline{P}_{\mathrm{cool}}, \eta_{\mathrm{heat}}\overline{P}_{\mathrm{heat}}]$.
\end{remark}

To summarize, given the thermal-NDC model, the state of the entire battery system at time instance $t$ is defined as
\begin{equation*}
\boldsymbol{x}(t) \triangleq \matt{V_{\mathrm{b}}(t) & V_{\mathrm{s}}(t) & T_{\mathrm{core}}(t) & T_{\mathrm{surf}}(t)}^{\top} \in \mathbb{R}^4.
\end{equation*}
The input of the system at time instance $t$ is defined as
\begin{equation*}
\boldsymbol{u}(t) \triangleq \matt{I(t) & P_{\mathrm{act}}(t)}^{\top} \in \mathbb{R}^2.
\end{equation*}
The output of the system at time instance $t$ is defined as
\begin{equation*}
\boldsymbol{y}(t) \triangleq \matt{ T_{\mathrm{surf}}(t) & V(t)}^{\top} \in \mathbb{R}^2.
\end{equation*}
The continuous-time system dynamics can be written as
\begin{equation} \label{eq:system_dynamics}
\begin{split}
\dot{\boldsymbol{x}} &= \boldsymbol{f}_{\mathrm{c}}(\boldsymbol{x}, \boldsymbol{u}), \\
\boldsymbol{y} &= \boldsymbol{g}(\boldsymbol{x}, \boldsymbol{u}),
\end{split}
\end{equation}
where the mapping $\boldsymbol{f}_{\mathrm{c}}: \mathbb{R}^4 \times \mathbb{R}^2 \mapsto \mathbb{R}^4$ and $\boldsymbol{g}: \mathbb{R}^4 \times \mathbb{R}^2 \mapsto \mathbb{R}^2$ are nonlinear and given by the combination of \eqref{eq:NDC_model} - \eqref{eq:heat_active}.




\section{Model Predictive Control Formulation} \label{sec:problem}

The problem of interest is to design a discrete-time control law for optimal charging while considering some safety and health constraints on LiBs, even if the ambient temperature is extremely high or low.
This section formulates a battery charging problem with a discrete-time MPC strategy.

\subsection{Constraints} \label{subsec:constraints}
This subsection first introduces some constraints which ensure safe and health-conscious charging.
To begin with, the SoC must be constrained to avoid overcharging, i.e.
\begin{equation} \label{constraint:soc}
\underline{\mathrm{SoC}} \leq \mathrm{SoC}(t) \leq \overline{\mathrm{SoC}}, \ \forall t.
\end{equation}
The current, terminal voltage and temperature must be within limits, i.e., $\forall t$,
\begin{subequations} \label{constraint:current_voltage_temp}
\begin{align}
\underline{I} &\leq I(t) \leq \overline{I}, \label{constraint:current_voltage_temp:1} \\
\underline{V} &\leq V(t) \leq \overline{V}, \label{constraint:current_voltage_temp:2} \\
\underline{T}_{\mathrm{core}} &\leq T_{\mathrm{core}}(t) \leq \overline{T}_{\mathrm{core}}. \label{constraint:current_voltage_temp:3}
\end{align}
\end{subequations}
The charging power $I(t)V(t)$ can also be bounded by the battery charger's power rating.
This paper omits the charging power constraint since both $I(t)$ and $V(t)$ are bounded.
The surface temperature $T_{\mathrm{surf}}$ largely depends on the ambient temperature $T_{\mathrm{amb}}$, whereas the core temperature $T_{\mathrm{core}}$ has much more impacts on the internal charging and discharging processes.
$T_{\mathrm{surf}}$ can also be bounded by critical temperatures at which surface thermal stress could cause significant damage to the battery \cite{zeng2021review}.
Nevertheless, it is reasonable to apply the bounds of $T_{\mathrm{core}}$ to $T_{\mathrm{surf}}$.

In the thermal-NDC model, $V_{\mathrm{b}}$ and $V_{\mathrm{s}}$ serve as an analogy to the Li-ion concentrations at the bulk inner and surface region of the electrode, respectively. Hence, the $V_{\mathrm{b}}$ and $V_{\mathrm{s}}$ also need constraints, as suggested in \cite{tian2020real}. Since $V_{\mathrm{b}} \leq V_{\mathrm{s}}$ during charging, one only needs to limit $V_{\mathrm{s}}$, i.e.
\begin{equation} \label{constraint:V_s}
\underline{V}_{\mathrm{s}} \leq V_{\mathrm{s}}(t) \leq \overline{V}_{\mathrm{s}}, \ \forall t.
\end{equation}
$V_{\mathrm{s}}-V_{\mathrm{b}}$ represents the Li-ion concentration gradient within the electrode \cite{fang2016health}.
This gradient indicates the undesired Lithium plating, a major factor of battery degradation in capacity and cycle life \cite{fang2016health,perez2017optimaljes,yin2021optimal,storch2021temperature,goshtasbi2024evtol}.
Consequently, it is crucial to limit the Li-ion concentration gradient to prevent Lithium plating \cite{fang2016health,storch2021temperature,goshtasbi2024evtol}.
Furthermore, this restriction should become stricter as SoC increases, as the battery becomes more vulnerable to a large Li-ion concentration gradient at higher SoC levels \cite{tian2020real}.
Therefore, this constraint can be written as
\begin{equation} \label{constraint:V_s_minus_V_b_old}
V_{\mathrm{s}}(t) - V_{\mathrm{b}}(t) \leq \beta_1 \mathrm{SoC}(t) + \beta_2, \ \forall t,
\end{equation}
where $\beta_1, \beta_2 \in \mathbb{R}$ are coefficients. Together with \eqref{eq:soc_define}, this constraint can be rewritten as
\begin{equation} \label{constraint:V_s_minus_V_b}
\begin{split}
\zeta(t) \triangleq &-\frac{C_{\mathrm{b}} \overline{V}_{\mathrm{b}} +C_{\mathrm{s}} \overline{V}_{\mathrm{s}}+ C_{\mathrm{b}}\beta_1}{C_{\mathrm{b}}\overline{V}_{\mathrm{b}}+C_{\mathrm{s}}\overline{V}_{\mathrm{s}}}V_{\mathrm{b}}(t) + \\
& \frac{C_{\mathrm{b}}\overline{V}_{\mathrm{b}}+C_{\mathrm{s}}\overline{V}_{\mathrm{s}}- C_{\mathrm{s}}\beta_1}{C_{\mathrm{b}}\overline{V}_{\mathrm{b}}+C_{\mathrm{s}}\overline{V}_{\mathrm{s}}}V_{\mathrm{s}}(t) \leq \beta_2, \ \forall t.
\end{split}
\end{equation}
\begin{remark} \label{remark:li_plating}
Lithium plating refers to the deposition of metallic Lithium on the surface of anode graphite particles \cite{goshtasbi2024evtol}.
One of the most important contributing factors to lithium plating is lithium-ion concentration gradients, as revealed in various studies \cite{perez2017optimaljes,yin2021optimal,storch2021temperature,goshtasbi2024evtol}. 
From the appendix of reference \cite{fang2016health}, $V_{\mathrm{s}} - V_{\mathrm{b}}$ plays a role similar to the Li-ion concentration gradient. Therefore, the constraint \eqref{constraint:V_s_minus_V_b_old} or \eqref{constraint:V_s_minus_V_b} prevents lithium plating during charging.
Regarding the parameters of this constraint, the values of $\beta_1$ and $\beta_2$ are typically determined by the degree to which this constraint becomes stricter as SoC increases.
These values can also be determined empirically from the anode plating overpotential trajectory over time, using experimental data obtained under various temperatures and charging current conditions \cite{romagnoli2019feedback,couto2021faster,goshtasbi2024evtol,mohtat2021algorithmic}.
\end{remark}

Finally, a constraint on the active battery temperature regulation system is considered, i.e.
\begin{equation} \label{constraint:heat_cool_power}
\underline{P}_{\mathrm{act}} \leq P_{\mathrm{act}}(t) \leq \overline{P}_{\mathrm{act}}, \ \forall t,
\end{equation}
where $\underline{P}_{\mathrm{act}}$ and $\overline{P}_{\mathrm{act}}$ indicate the battery cooling and heating power limits, respectively.

\subsection{State-Feedback Model Predictive Control}

This subsection introduces the formulation of a discrete-time state-feedback MPC for the battery charging problem.
Let $N \in \mathbb{N}$ denote the number of steps in a prediction horizon and $\boldsymbol{x}_k \coloneqq \boldsymbol{x}(t_k)$.
By Euler integration with a planning time interval $\Delta_{\mathrm{p}} > 0$, one obtains the following discretization of the continuous system in \eqref{eq:system_dynamics} for MPC:
\begin{equation} \label{eq:system_dynamics_discrete}
\begin{split}
\boldsymbol{x}_{k+1}& = \boldsymbol{x}_{k} + \Delta_{\mathrm{p}} \boldsymbol{f}_{\mathrm{c}}(\boldsymbol{x}_k, \boldsymbol{u}_k), \\
\boldsymbol{y}_k &= \boldsymbol{g}(\boldsymbol{x}_k, \boldsymbol{u}_k). \\
\end{split}
\end{equation}
Denote $\boldsymbol{x}_{0:N|k} \triangleq \text{col}\{\boldsymbol{x}_{k},\boldsymbol{x}_{k+1|k}\cdots,\boldsymbol{x}_{k+N|k}\} \in \mathbb{R}^{4(N+1)}$ the state at time $t_k$ and the states from the future time $t_{k+1}$ to $t_{k+N}$ that are predicted at time $t_k$; similarly $\boldsymbol{u}_{0:N-1|k} \triangleq \text{col}\{\boldsymbol{u}_{k|k},\cdots,\boldsymbol{u}_{k+N-1|k}\} \in \mathbb{R}^{2N}$;
particularly denote $\mathrm{SoC}_{0:N|k} \triangleq \text{col}\{\mathrm{SoC}_{k|k},\cdots,\mathrm{SoC}_{k+N|k}\} \in \mathbb{R}^{N+1}$ and $V_{0:N-1|k} \triangleq \text{col}\{V_{k|k},\cdots,V_{k+N-1|k}\} \in \mathbb{R}^{N}$.
Define the objective function $J \coloneqq J(\boldsymbol{x}_{0:N|k},\boldsymbol{u}_{0:N-1|k})$ as
\begin{equation} \label{eq:define_J}
\begin{split}
J \triangleq &w_1\textstyle\sum_{j=0}^{N} (\mathrm{SoC}_{k+j|k}-\mathrm{SoC}_{\mathrm{r}})^2 + \\
& w_2\textstyle\sum_{j=0}^{N-2} (I_{k+j+1|k}-I_{k+j|k})^2 + \\
& w_3\textstyle\sum_{j=0}^{N-2} (P_{\mathrm{act},k+j+1|k}-P_{\mathrm{act},k+j|k})^2,
\end{split}
\end{equation}
where $\mathrm{SoC}_{\mathrm{r}}$ is the target SoC to be charged, $w_1, w_2, w_3>0$ are some weights.
The first term of this objective function reflects that the battery is desired to be charged as soon as possible. The second and third terms encourage the current's and thermal power's smoothness over time, respectively.

\begin{remark} \label{remark:control_smoothness}
The required smoothness of the charging current imposed by \eqref{eq:define_J} may vary or even be unnecessary, depending on the specifications of the power electronics.
For tutorial purposes, this paper adopts a general form to define the smoothness requirement as part of the objective, but it can be adjusted to meet specific needs.
The smoothness requirement for $P_{\mathrm{act}}$ in \eqref{eq:define_J} is imposed for similar reasons.
\end{remark}

Then at time $t_k$, given the state $\boldsymbol{x}_{k}$ and the ambient temperature $T_{\mathrm{amb},k} \coloneqq T_{\mathrm{amb}}(t_k)$, assuming that $T_{\mathrm{amb},k+j|k} \equiv T_{\mathrm{amb},k}, \  \forall j \in \llbracket 0,N \rrbracket$, the optimal control can be determined by
\begin{mini}|s|
{ \boldsymbol{u}_{0:N-1|k} }{ J(\boldsymbol{x}_{0:N|k},\boldsymbol{u}_{0:N-1|k}) \label{eq:oc_mpc_1}}
{}{}
\addConstraint{ \boldsymbol{x}_{k+j+1|k} = \boldsymbol{x}_{k+j|k} + \Delta_{\mathrm{p}} \boldsymbol{f}_{\mathrm{c}}(\boldsymbol{x}_{k+j|k}, \boldsymbol{u}_{k+j|k})}
\addConstraint{\boldsymbol{y}_{k+j|k}= \boldsymbol{g}(\boldsymbol{x}_{k+j|k},\boldsymbol{u}_{k+j|k})}
\addConstraint{\forall j \in \llbracket 0,N-1 \rrbracket \  \text{with given } \boldsymbol{x}_{k}}
\addConstraint{\text{constraints } \eqref{constraint:soc}, \eqref{constraint:current_voltage_temp:3}-\eqref{constraint:heat_cool_power}, \ \forall j \in \llbracket 0,N \rrbracket}
\addConstraint{\text{constraints } \eqref{constraint:current_voltage_temp:1}, \eqref{constraint:current_voltage_temp:2}, \ \forall j \in \llbracket 0,N-1 \rrbracket.}
\end{mini}

The state-feedback MPC algorithm is summarized in Algorithm~\ref{alg:mpc_state}.
As shown by Line 2, the system state is observed with a smaller sampling time $\Delta_{\mathrm{s}}$, i.e., $\Delta_{\mathrm{s}} < \Delta_{\mathrm{p}}$.
The MPC is computed for every time interval $\Delta_{\mathrm{p}}$ given the present state $\boldsymbol{x}_k$ at time $t_k$, as shown by Line 3.
Then for every time instance $t_i$, the input $\boldsymbol{u}_i$ is calculated by interpolating the optimal control trajectory from the MPC with zero-order hold, as indicated in Line 8.

\begin{algorithm}
\caption{State-Feedback MPC}\label{alg:mpc_state}
\DontPrintSemicolon
\KwIn{$k = -1$, $i = 0$, $t_0=0$, $\Delta_{\mathrm{p}}$, $\Delta_{\mathrm{s}}$}
\While {$\mathrm{SoC}_i < \mathrm{SoC}_{\mathrm{r}}$} {
observe $\boldsymbol{x}_i$ and $T_{\mathrm{amb},i}$ at time $t_i$\;

\uIf(\tcp*[h]{for every time interval $\Delta_{\mathrm{p}}$}){$t_i \ \% \ \Delta_{\mathrm{p}} == 0$}
{
$k \gets k + 1$\;
$\boldsymbol{u}^*_{0:N-1|k} \gets$  solve $\eqref{eq:oc_mpc_1}$ given $\boldsymbol{x}_k$ at time $t_k$ and $T_{\mathrm{amb},k+j|k} \equiv T_{\mathrm{amb},k}, \  \forall j \in \llbracket 0,N \rrbracket$\;
$t_k \gets t_k + \Delta_{\mathrm{p}}$
\;}
\Else{
perform $\boldsymbol{u}_i$ by interpolating $\boldsymbol{u}^*_{0:N-1|k}$ with present time $t_i$ and zero-order hold\;
}
$t_{i} \gets t_i + \Delta_{\mathrm{s}}$, $i \gets i+1$\;
}
\end{algorithm}


\subsection{System Parameters and Identification} \label{subsec:system_para}


The necessary parameters for the battery dynamics \eqref{eq:system_dynamics} and the constraints \eqref{constraint:soc} - \eqref{constraint:heat_cool_power} are given by Table~\ref{table:parameter}. The system parameters are given in \cite{tian2020nonlinear}, using a 3 Ah Panasonic NCR-18650B LiB battery. The constraint parameters are derived from \cite{tian2020real}. $\overline{P}_{\mathrm{heat}}$ and $\overline{P}_{\mathrm{cool}}$ are determined by scaling the power limits in \cite{hamednia2023optimal}.
The mapping $h(\cdot)$ in \eqref{eq:NDC_model:one_terminal_v} is parameterized by a fifth-order polynomial \cite{tian2020nonlinear}, i.e.
\begin{equation} \label{eq:h_mapping}
h(V_{\mathrm{s}}) = \textstyle\sum_{i=0}^5 \alpha_i V_{\mathrm{s}}^i.
\end{equation}
The model parameters are validated for charging current up to 4.5 A, whose maximum charging C-rate is 1.5C.
This paper defines the upper bound for charging current as $\overline{I}=3$ A.
For the sake of readability, the rest of this paper adopts degree Celsius to express the same temperature in Kelvin.

To identify and calibrate the battery parameters, a two-step procedure is adopted as described in \cite{tian2020nonlinear}.
This process involves initially identifying the $h(\cdot)$ mapping and subsequently estimating the impedance and capacitance parameters.

First, the SoC-OCV relationship of the proposed model is given by $\text{OCV} = h(\text{SoC})$.
The $h(\cdot)$ mapping can be identified by fitting it with a battery's SoC-OCV data.
To obtain this data, the battery is discharged at a low current (e.g., 1/25 C-rate) from full to empty.
During this process, the terminal voltage $V$ is taken as OCV, and the SoC is calculated using the Coulomb counting method \cite{tian2020nonlinear}.
The mapping $h(\cdot)$ is then identified as a parameterized polynomial by solving a least squares problem.

Second, to identify the impedance and capacitance parameters, the battery is discharged at a constant current of normal magnitude.
These parameters are estimated by expressing the terminal voltage in terms of the parameters and fitting this expression to the measurement data \cite{tian2020nonlinear}.
For further details, refer to \cite{tian2020nonlinear,tu2024system}.

\begin{table}
\centering
\begin{threeparttable}
\caption{Battery Parameters} \label{table:parameter}
\begin{tabular}{c c c c c c}
\toprule
Parameter & Value & Param. & Value & Param. & Value \\
\midrule
$C_{\mathrm{b}}$ & 10037 F & $C_{\mathrm{core}}$ & 40 J/K & $\underline{\mathrm{SoC}}$ & 0\%\\
$C_{\mathrm{s}}$ & 973 F & $C_{\mathrm{surf}}$ & 10 J/K & $\overline{\mathrm{SoC}}$ & 100\% \\
$R_{\mathrm{b}}$ & 0.019 $\Omega$ & $R_{\mathrm{core}}$ & 4 K/W & $\underline{I}$ & 0 A \\
$\gamma_1$ & 0.026 $\Omega$ & $R_{\mathrm{surf}}$ & 7 K/W & $\overline{I}$ & 3 A\\
$\gamma_2$ & 0.061 $\Omega$ & $\kappa_1$ & 30 K & $\underline{V}$ & 0 V \\
$\gamma_3$ & 14.36 & $\kappa_2$ & 70 K & $\overline{V}$ & 4.2 V\\
$\alpha_0$ & 3.2 V & $T_{\mathrm{ref}}$ & 298.15 K & $\underline{T}_{\mathrm{core}}$ & 263.15 K \\
$\alpha_1$ & 2.59 & $\eta_{\mathrm{act}}$ & 87\% & $\overline{T}_{\mathrm{core}}$ & 328.15 K \\
$\alpha_2$ & -9.003 V$^{-1}$ & $\beta_1$ & -0.04 V & $\underline{V}_{\mathrm{b}}$ & 0 V \\
$\alpha_3$ & 18.87 V$^{-2}$ & $\beta_2$ & 0.08 V& $\overline{V}_{\mathrm{b}}$ & 1 V \\
$\alpha_4$ & -17.82 V$^{-3}$ & $\underline{P}_{\mathrm{act}}$ & -8 W & $\underline{V}_{\mathrm{s}}$ & 0 V \\
$\alpha_5$ & 6.325 V$^{-4}$ & $\overline{P}_{\mathrm{act}}$ & 8 W & $\overline{V}_{\mathrm{s}}$ & 1 V \\
\bottomrule
\end{tabular}
\end{threeparttable}
\centering
\end{table}

\section{Numerical Experiments} \label{sec:num_sim}

This section presents several case studies with numerical experiments to illustrate the effectiveness of the proposed thermal-NDC model in integrated charging and active thermal management under extreme ambient temperatures. Additionally, this section investigates the impact of MPC parameters, battery core temperature, and active thermal power on fast charging.
Several insights are revealed and summarized in this section to explain why the decisions made by the proposed algorithm lead to energy-efficient and health-aware optimal fast charging.

\subsection{Basic Case Study} \label{subsec:basic_case_study}

With the proposed thermal-NDC model, this subsection performs several case studies with low, mild, and high ambient temperatures and compares the battery charging performance with several strategies to verify the effectiveness of the proposed charging strategies.
The list of all strategies is summarized in Table \ref{table:strategy_list_basic}. Strategy B adopts the same MPC scheme \eqref{eq:oc_mpc_1} without active thermal management involved in the MPC formulation to mimic the existing battery modeling without external thermal influence. B utilizes a separate PID temperature controller, which is common in industry, i.e.
\begin{equation} \label{eq:temp_P_ctrl}
\begin{split}
e(t_k) &\coloneqq e_k = T_{\mathrm{core,r}} - T_{\mathrm{core},k}, \\
P_{\mathrm{act},k} &= [K_{\mathrm{P}} e_k + K_{\mathrm{I}} \textstyle\sum_{j=0}^k e_j + K_{\mathrm{D}} \dot{e}(t_k)]_{\underline{P}_{\mathrm{act}}}^{\overline{P}_{\mathrm{act}}}, \\
\end{split}
\end{equation}
where $T_{\mathrm{core,r}}$ is the desired battery core temperature; $T_{\mathrm{core,k}}$ is the battery core temperature at time $t_k$; $K_{\mathrm{P}}, K_{\mathrm{I}}, K_{\mathrm{D}} \geq 0$ are the proportional, integral, and derivative gain, respectively; $[\cdot]_{\underline{P}_{\mathrm{act}}}^{\overline{P}_{\mathrm{act}}}$ denotes the clip function with maximum value $\overline{P}_{\mathrm{act}}$ and minimum value $\underline{P}_{\mathrm{act}}$; $\dot{e}(t_k) \coloneqq -\dot{T}_{\mathrm{core}}(t_k)$, and $\dot{T}_{\mathrm{core}}(t_k)$ is estimated by applying the optimal control $\boldsymbol{u}^*_{k|k}$ at time $t_k$ without active thermal control to the system dynamics \eqref{eq:lumped_thermal_model}.

\begin{table}
\centering
\begin{threeparttable}
\caption{Strategy Description for Basic Case Study} \label{table:strategy_list_basic}
\begin{tabularx}{\linewidth}{l X}
\toprule
\multicolumn{1}{c}{Name} & \multicolumn{1}{c}{Description} \\
\midrule
P & the proposed MPC \eqref{eq:oc_mpc_1}, optimization initial guess given by forward propagating dynamics with zero inputs \\
\midrule
P1 & same as P, optimization initial guess given by forward propagating dynamics with maximum charging current and $P_{\mathrm{act},k}$; $P_{\mathrm{act},k}$ is forwardly calculated by \eqref{eq:temp_P_ctrl}  with $T_{\mathrm{core,r}}=45$ $^{\circ}\text{C}$ \\
\midrule
A & the proposed MPC \eqref{eq:oc_mpc_1} without active thermal management, i.e., $\underline{P}_{\mathrm{act}} = \overline{P}_{\mathrm{act}} = 0$ W, $P_{\mathrm{act}}(t)=0, \  \forall t$ \\
\midrule
B & A, battery temperature is separately controlled by the PID controller \eqref{eq:temp_P_ctrl}, $T_{\mathrm{core,r}}=25$ $^{\circ}\text{C}$ \\
\midrule
C & same as B, $T_{\mathrm{core,r}}=35$ $^{\circ}\text{C}$ \\
\midrule
D & same as B, $T_{\mathrm{core,r}}=45$ $^{\circ}\text{C}$ \\
\midrule
E & same as B, $T_{\mathrm{core,r}}=50$ $^{\circ}\text{C}$ \\
\bottomrule
\end{tabularx}
\end{threeparttable}
\centering
\end{table}

The parameters used in the basic case study are:
$\mathrm{SoC_r} = 0.90$, $w_1 = 40$, $w_2 = 0.1$, $w_3 = 0.1$, $\Delta_{\mathrm{p}} = 5$ s, $\Delta_{\mathrm{s}} = 1$ s, $N=40$, $K_{\mathrm{P}}=0.5$, $K_{\mathrm{I}}=0.01$, $K_{\mathrm{P}}=150$.
The discrete-time system dynamics in the simulation is updated in every time interval $\Delta_{\mathrm{s}}$ and given by
\begin{equation}
\boldsymbol{x}_{i+1} =\boldsymbol{x}_{i} + \Delta_{\mathrm{s}} \boldsymbol{f}_{\mathrm{c}}(\boldsymbol{x}_i, \boldsymbol{u}_i).
\end{equation}
The PID gains are tuned appropriately to minimize the rising time, overshoot, and steady-state error on temperature tracking.
When strategies involve solving the optimal control instance \eqref{eq:oc_mpc_1}, the optimal control instance is programmed by Python with CasADi \cite{andersson2019casadi}, compiled as a C library with a nonlinear programming solver IPOPT \cite{wachter2006implementation} with MUMPS, and then executed at run-time in Python.
All the simulations are run on a 2017 MacBook Pro equipped with a 3.1-GHz Inter Core i7 and 16 GB RAM.

\begin{figure*}
\subfloat[Result by P and A with $T_{\mathrm{amb}} = 25$ $^{\circ}\text{C}$.]
{\label{fig:temp_result:mild} \includegraphics[width=0.32\linewidth]{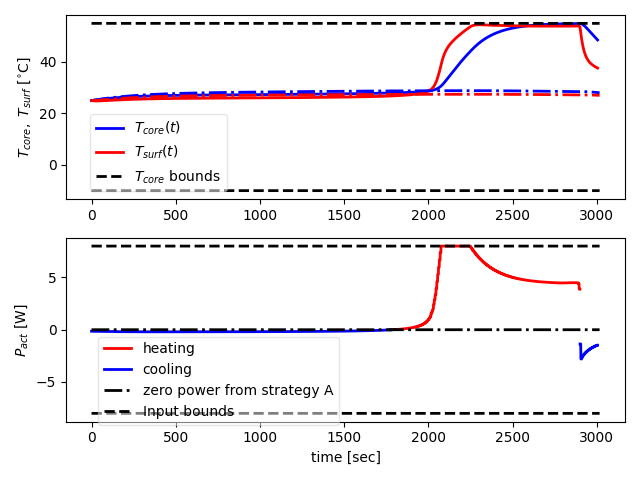}}
\hfill
\subfloat[Result by P with $T_{\mathrm{amb}} = 70$ $^{\circ}\text{C}$.]
{\label{fig:temp_result:high} \includegraphics[width=0.32\linewidth]{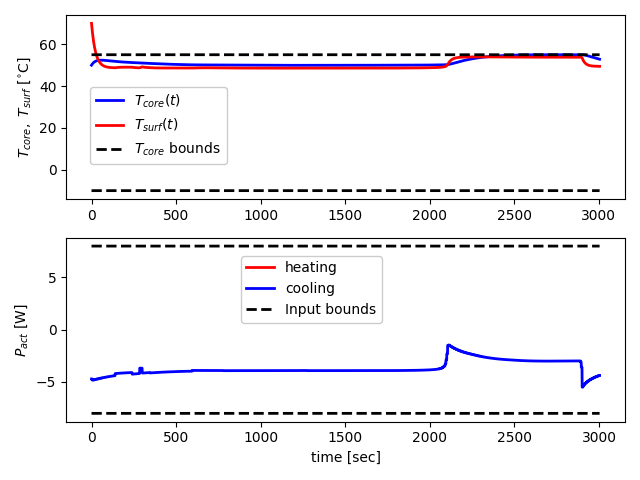}}
\hfill
\subfloat[Result by P with $T_{\mathrm{amb}} = -25$ $^{\circ}\text{C}$.]
{\label{fig:temp_result:low} \includegraphics[width=0.32\linewidth]{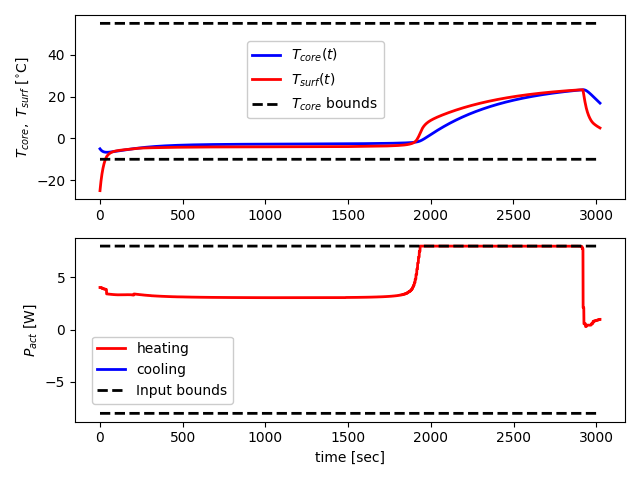}}
\caption{Battery core and surface temperature and active thermal power by some strategies in different ambient temperatures. In the upper portion of each subfigure, the solid lines in blue and red represent $T_{\mathrm{core}}$ and $T_{\mathrm{surf}}$ by strategy P. In the lower portion of each subfigure, the solid lines in blue and red represent the active cooling and heating power by P. In subfigure (a), the dash-dot lines in blue and red represent $T_{\mathrm{core}}$ and $T_{\mathrm{surf}}$ by strategy A; the dash-dot line in black represents the zero active thermal power by A. Note that there is no constraint on $T_{\mathrm{surf}}$.} \label{fig:temp_result}
\end{figure*}

\subsubsection{Mild Ambient Temperature}
The ambient temperature $T_{\mathrm{amb}} = 25$ $^{\circ}\text{C}$. The initial state $\boldsymbol{x}_0 = \matt{0.1 \text{ V} & 0.1 \text{ V} & 25\,^{\circ}\text{C} & 25\,^{\circ}\text{C}}^{\top}$.
The simulation results are shown in Table \ref{table:basic_mild}, where $T_{\mathrm{chg}}$ indicates the total charging time; $T_{\mathrm{comp}}$ indicates the average computational time and its standard deviation; Energy indicates the total energy used for charging and thermal management; Efficiency indicates the charging efficiency (the ratio of the energy used for raising SoC to the total energy), i.e., $$\frac{\sum_{k=0}^{T_{\mathrm{chg}}/\Delta_{\mathrm{s}}} I_k \cdot h(\mathrm{SoC}_k) \Delta_{\mathrm{s}}}{\sum_{k=0}^{T_{\mathrm{chg}}/\Delta_{\mathrm{s}}} (I_kV_k+|P_{\mathrm{act},k}|) \Delta_{\mathrm{s}}}.$$

According to Table \ref{table:basic_mild}, strategies P and P1 have the smallest charging time. As for total consumed energy and efficiency, A outperforms the others since charging the battery without active thermal management will not violate the battery temperature constraint when the ambient temperature is mild.
Fig. \ref{fig:temp_result:mild} illustrates the trajectories of battery core and surface temperature and active thermal power for strategies P and A, where strategy P chooses to warm the battery after about 1800 s and thus results in faster charging.
From strategies B to D, as the desired battery core temperature $T_{\mathrm{core,r}}$ increasing from 25 $^{\circ}\text{C}$ to 45 $^{\circ}\text{C}$, the charging time decreases from 3019 s to 3009 s. But when the desired core temperature $T_{\mathrm{core,r}}=50$ $^{\circ}\text{C}$ is closed to the upper bound $\overline{T}_{\mathrm{core}} = 55$ $^{\circ}$C, the PID controller violates $\overline{T}_{\mathrm{core}}$ at 690 s - 1090 s since it has no predictive information on how the system dynamics will evolve. These observations are consistent with the statement that heating the battery could benefit the charging speed \cite{wang2016lithium,yang2018fast}.


\begin{table}
\centering
\begin{threeparttable}
\caption{Battery Charging Result in Mild Ambient Temperature} \label{table:basic_mild}
\begin{tabular}{c c c c c}
\toprule
Strategy & $T_{\mathrm{chg}}$ [s] & Energy [kJ] & Efficiency & $T_{\mathrm{comp}}$ [ms] \\
\midrule
P & \bf{3005} & 38.98 & 83.10\% & $23.59\pm4.38$\\
P1 & \bf{3005} & 38.99 & 83.08\% & $21.66\pm4.99$\\
A & 3017 & \bf{33.42} & \bf{96.93\%} & $22.14\pm3.67$\\
B & 3019 & 35.25 & 91.90\% & $21.47\pm2.16$\\
C & 3013 & 37.24 & 86.98\% & $21.90\pm2.36$\\
D & 3009 & 42.88 & 75.54\% & $21.53\pm2.38$\\
E$^{\dagger}$ & 3416 & 47.39 & 68.35\% & $23.40\pm6.58$\\
\bottomrule
\end{tabular}
\begin{tablenotes}
\small
\item[$\dagger$] Infeasible at 690 - 1090 s; violates $\overline{T}_{\mathrm{core}}$
\end{tablenotes}
\end{threeparttable}
\centering
\end{table}


\subsubsection{High Ambient Temperature} \label{subsubsec:high_temp}
The ambient temperature $T_{\mathrm{amb}} = 70$ $^{\circ}\text{C}$. The initial state $\boldsymbol{x}_0 = \matt{0.1 \text{ V} & 0.1 \text{ V} & 50\,^{\circ}\text{C} & 70\,^{\circ}\text{C}}^{\top}$.
Note that this initial state is more challenging than $T_{\mathrm{core}} = T_{\mathrm{surf}} = 50$ $^{\circ}\text{C}$ since the heat convection from the surface to the core could further increase the core temperature.
The simulation results are shown in Table \ref{table:basic_high}, where strategies P and P1 have the smallest charging time but have slightly different total consumed energy and efficiency. A conjecture is that the real difference in charging time could be less than the time resolution of the simulation $\Delta_{\mathrm{s}} = 1$ s; the slightly different control trajectories lead to different values on the accumulated energy consumption and efficiency.
Strategies A - E cannot find a feasible solution for solving MPC instances over some duration when the ambient temperature is high, which reflects that these strategies are not applicable at run-time in this scenario.
A practical workaround for these strategies is to solely adjust the battery temperature to a mild range first, then charge the battery. This solution leads to longer charging time because it cannot jointly determine the battery's electrical and thermal control.
This case study verifies the effectiveness of the proposed strategies P and P1 with high ambient temperature.
Fig. \ref{fig:temp_result:high} illustrates the trajectories of battery temperature and active thermal power for strategy P, where the active cooling power is regulated to balance the heat convention among the ambient, the battery surface, and the battery core.

\begin{table}
\centering
\begin{threeparttable}
\caption{Battery Charging Result in High Ambient Temperature} \label{table:basic_high}
\begin{tabular}{c c c c c}
\toprule
Strategy & $T_{\mathrm{chg}}$ [s] & Energy [kJ] & Efficiency & $T_{\mathrm{comp}}$ [ms] \\
\midrule
P & \bf{3004} & \bf{44.43} & \bf{72.91\%} & $25.45\pm4.85$\\
P1 & \bf{3004} & 44.45 & 72.87\% & $20.88\pm4.99$\\
A$^{\dagger}$ & N/A & N/A & N/A & $49.76\pm4.28$\\
B$^{\ddagger}$ & 3098 & 58.20 & 55.65\% & $21.52\pm4.17$\\
C$^{*}$ & 3131 & 55.24 & 59.72\% & $21.70\pm5.00$\\
D$^{**}$ & 3416 & 49.55 & 65.37\% & $23.74\pm7.76$\\
E$^{\dagger}$ & N/A & N/A & N/A & $42.04\pm2.96$\\
\bottomrule
\end{tabular}
\begin{tablenotes}
\small
\item[$\dagger$] Infeasible until simulation timeouts
\item[$\ddagger$] Infeasible in the first 70 s
\item[$*$] Infeasible in the first 105 s
\item[$**$] Infeasible in the first 385 s
\end{tablenotes}
\end{threeparttable}
\centering
\end{table}

\subsubsection{Low Ambient Temperature}
The ambient temperature $T_{\mathrm{amb}} = -25$ $^{\circ}\text{C}$. The initial state $\boldsymbol{x}_0 = \matt{0.1 \text{ V} & 0.1 \text{ V} & -5\,^{\circ}\text{C} & -25\,^{\circ}\text{C}}^{\top}$.
The simulation results are shown in Table \ref{table:basic_low}.
Considering the constraint satisfaction and the MPC solution feasibility, strategy P outperforms the others in charging time, total energy consumption, and efficiency. Strategy P1 has the same charging time as P but the energy consumption and efficiency are slightly different than P1's due to the same reason mentioned in Section \ref{subsubsec:high_temp}. Strategies A - E cannot find a feasible solution for solving MPC instances over some duration since the battery temperature is close to the lower bound.
Since the constraints on $T_{\mathrm{core}}$, $V$, and $\zeta$ limit the charging speed and strategies B - E violate some of them, their total charging time are less than P's and P1's.
Violating these constraints during fast charging could jeopardize the battery health \cite{tian2020real}.
Thus, these strategies are not applicable at run-time when the ambient temperature is low.
Fig. \ref{fig:temp_result:low} illustrates the trajectories of battery temperature and active thermal power for strategy P, where the active heating power hits the upper bound to warm up the battery for fast charging.

\begin{table}
\centering
\begin{threeparttable}
\caption{Battery Charging Result in Low Ambient Temperature} \label{table:basic_low}
\begin{tabular}{c c c c c}
\toprule
Strategy & $T_{\mathrm{chg}}$ [s] & Energy [kJ] & Efficiency & $T_{\mathrm{comp}}$ [ms] \\
\midrule
P & \bf{3023} & \bf{47.63} & 68.01\% & $24.67\pm4.85$\\
P1 & \bf{3023} & 47.71 & \bf{67.89\%} & $21.91\pm5.45$\\
A$^{\dagger}$ & N/A & N/A & N/A & $106.88\pm63.50$\\
B$^{\ddagger}$ & 3022 & 57.60 & 56.24\% & $21.93\pm9.20$\\
C$^{\ddagger\mathsection}$ & 3022 & 57.60 & 56.24\% & $21.96\pm9.05$\\
D$^{\ddagger\mathsection}$ & 3023 & 47.63 & 68.01\% & $24.67\pm4.85$\\
E$^{\ddagger\mathsection}$ & 3022 & 57.60 & 56.24\% & $21.98\pm9.07$\\
\bottomrule
\end{tabular}
\begin{tablenotes}
\small
\item[$\dagger$] Infeasible until simulation timeouts
\item[$\ddagger$] Infeasible in the first 25 s
\item[$\mathsection$] Unable to reach $T_{\mathrm{core,r}}$ due to power limits
\end{tablenotes}
\end{threeparttable}
\centering
\end{table}

\subsubsection{Summary on Basic Case Study}
To summarize the three basic case studies above, the proposed strategies P and P1 outperform the others in charging time given three different ambient temperatures, which cover most of the battery operational conditions.
Strategies A - E are not applicable at run-time when the ambient temperature is extreme since they might not find a feasible solution.
Strategy P and P1 outperform the others in energy consumption and efficiency with both high and low ambient temperature even though these factors are not explicitly considered in the objective function.
This is because, with the proposed thermal-NDC model, P and P1 can jointly determine control such that the battery temperature and SoC can be mutually beneficial to each other.
To further improve energy efficiency and reduce energy consumption with mild ambient temperature, one could restrict the power limit on $P_{\mathrm{act}}$ given a certain range of mild ambient temperature.

\begin{figure}[ht]
\centering
\includegraphics[width=0.48\textwidth]{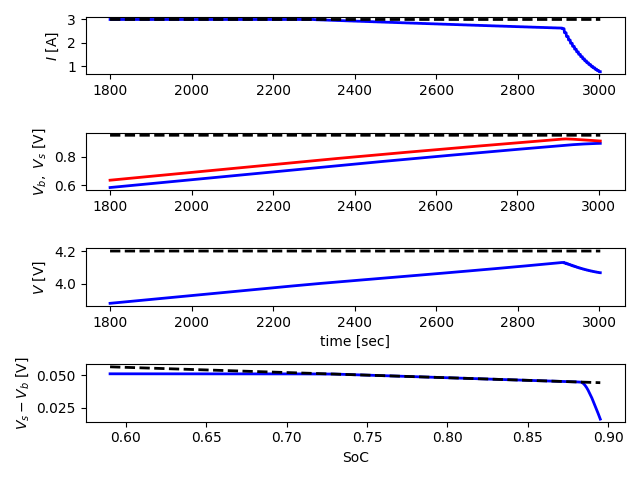}
\caption{A trajectory segment for $I$, $V_{\mathrm{b}}$, $V_{\mathrm{s}}$, $V$, and $V_{\mathrm{s}}(t) - V_{\mathrm{b}}(t)$ vs $\beta_1 \mathrm{SoC}(t) + \beta_2$ with $T_{\mathrm{amb}} = 70$ $^{\circ}\text{C}$. The red and blue curves in the second subfigure indicate $V_{\mathrm{s}}$ and $V_{\mathrm{b}}$, respectively. The black dashed lines in the top three subfigures are the upper bounds. The black dashed line in the last subfigure indicates $\beta_1 \mathrm{SoC}(t) + \beta_2$ as $\mathrm{SoC}(t)$ increases over time.}
\label{fig:extra}
\end{figure}

The advantage of the proposed thermal-NDC model can be also reflected by an observation from Fig.~\ref{fig:temp_result}, where $P_{\mathrm{act}}$ drops toward cooling to reduce $T_{\mathrm{core}}$ and $T_{\mathrm{surf}}$ at the end of charging.
In fact, as Fig.~\ref{fig:extra} illustrated, the charging current also drops at the same time.
The dropping on both $P_{\mathrm{act}}$ and $I$ is to reduce $V_{\mathrm{s}}$ and $V$ such that they will not violate the constraints as they approach the upper bounds closely. Also, as SoC increases, the upper bound on the Li-ion concentration gradient within the electrode, i.e., $\beta_1 \mathrm{SoC}(t) + \beta_2$, decreases. This again requires $V_{\mathrm{s}}$ to reduce such that $V_{\mathrm{s}} - V_{\mathrm{b}}$ will not violate its upper bound.
The system behaviors with $T_{\mathrm{amb}} = -25$ $^{\circ}\text{C}$ and $T_{\mathrm{amb}} = 25$ $^{\circ}\text{C}$ are similar to Fig.~\ref{fig:extra}, thereby omitted.

As for the computational time between P and P1, Tukey’s HSD (honestly significant difference) Test \cite[Chapter~14]{lowry2014concepts} is performed for three cases and it reveals a statistically significant difference between the computational time of strategies P and P1; the computational time of P is statistically greater than P1's with the p-value 0.00 $<$ 0.05.
This indicates that the initial guess of P1's MPC problem, i.e., regulating $T_{\mathrm{core}}$ to $T_{\mathrm{core,r}}=45$ $^{\circ}\text{C}$ in all three ambient temperatures, is closer to the optimum than P's initial guess, which benefits the run-time computation.

A similar conclusion can be drawn based on the observation from Fig.~\ref{fig:temp_result}, where the MPC strategy P seems to maintain a high $T_{\mathrm{core}}$ for fast charging.
These two observations are consistent with the conclusions made by \cite{wang2016lithium,yang2018fast}.
However, compared with the results in mild and high ambient temperatures, it is uncertain that raising the heating power bound could further speed up the charging given low ambient temperature.
Also, comparing the computational time of P and P1 in the low and high ambient temperatures, the initialization of P1, i.e., regulating $T_{\mathrm{core}}$ to $T_{\mathrm{core,r}}=45$ $^{\circ}\text{C}$, might be closer to the optimal solution than strategy P's initialization.
Thus, the next subsection investigates how battery core temperature and active heating would affect the charging speed, and verify whether an MPC strategy with a long enough horizon can obtain an optimal solution that implicitly encourages warming up the battery to an optimal temperature.

\subsection{Other Factors}

\begin{table}
\centering
\begin{threeparttable}
\caption{Strategy Description for Other Factors}
\label{table:strategy_list_basic_cont}
\begin{tabularx}{\linewidth}{l X}
\toprule
\multicolumn{1}{c}{Name} & \multicolumn{1}{c}{Description} \\
\midrule
P2 & same as P$^{\dagger}$, objective function using $J_2$ in \eqref{eq:objective_track_t_core}, $T_{\mathrm{core,r}}=45^{\circ}\text{C}$\\
\midrule
P3 & same as P2, $N=40$, $T_{\mathrm{core,r}}=55$ $^{\circ}\text{C}$\\
\midrule
P4 & same as P, $N=80$\\
\midrule
P5 & same as P, $N=120$\\
\bottomrule
\end{tabularx}
\begin{tablenotes}
\small
\item[$\dagger$] Note: for strategy P, $N=40$
\end{tablenotes}
\end{threeparttable}
\centering
\end{table}

According to \cite{wang2016lithium}, the Lithium-ion battery temperature plays a significant role during charging and discharging, especially when the ambient temperature is below 0 $^{\circ}\text{C}$.
Heating Lithium-ion batteries at the beginning of charging could improve the charging speed and capacity retention even if the ambient temperature is -30 $^{\circ}\text{C}$ \cite{wang2016lithium,yang2018fast}.
Therefore, with the proposed thermal-NDC model, this subsection performs several numerical experiments in low ambient temperatures to investigate how the MPC horizon and battery temperature affect the charging speed.

To investigate how $T_{\mathrm{core}}$ affects the charging speed, a revised objective function is given to explicitly warm up the battery core to a desired temperature while charging, i.e.
\begin{equation} \label{eq:objective_track_t_core}
\begin{split}
J_2 \triangleq &w_1\textstyle\sum_{j=0}^{N} (\mathrm{SoC}_{k+j|k}-\mathrm{SoC}_{\mathrm{r}})^2 + \\
&w_2\textstyle\sum_{j=0}^{N-2} (I_{k+j+1|k}-I_{k+j|k})^2 + \\
& w_3\textstyle\sum_{j=0}^{N-2} (P_{\mathrm{act},k+j+1|k}-P_{\mathrm{act},k+j|k})^2 + \\
& w_4\textstyle\sum_{j=0}^{N} (T_{\mathrm{core},k+j|k}-T_{\mathrm{core,r}})^2,
\end{split}
\end{equation}
where $T_{\mathrm{core,r}}$ is a prescribed target battery core temperature, $w_1, w_2, w_3, w_4 > 0$ are some weights. In this subsection, parameters are the same as the ones in Section \ref{subsec:basic_case_study}, except for explicit annotations. $w_4 = 0.5$ when applicable.
The power limits of $P_{\mathrm{act}}$ are relaxed as $[-24 \text{ W}, 24 \text{ W}]$ such that the system has enough power to regulate $T_{\mathrm{core}}$ within the range [-10 $^{\circ}$C, 55 $^{\circ}$C], in low ambient temperature.
Table~\ref{table:strategy_list_basic_cont} summarizes the additional strategies to be compared. The comparison results are summarized in Table \ref{table:higher_power_low_temp}.

Regarding the computational time of the proposed strategies P and P1 - P5, a one-way analysis of variance (one-way ANOVA) \cite[Chapter~14]{lowry2014concepts} is performed, which reveals a statistically significant difference in computational time between at least two groups with a p-value of $0.00 < 0.05$. Then Tukey’s HSD Test reveals a relation with statistical significance among these groups, i.e., P1 $<$ P $<$ P2 $<$ P3 $<$ P4 $<$ P5.


\begin{table}
\centering
\begin{threeparttable}
\caption{Battery Charging Result in Low Ambient Temperature with Higher Heating/Cooling Power Limits} \label{table:higher_power_low_temp}
\begin{tabular}{c c c c c}
\toprule
Strategy & $T_{\mathrm{chg}}$ [s] & Energy [kJ] & Efficiency & $T_{\mathrm{comp}}$ [ms] \\
\midrule
P & 3005 & \bf{59.68} & \bf{54.28\%} & $23.50\pm4.39$\\
P1 & 3005 & 59.71 & 54.24\% & $20.99\pm4.86$\\
P2 & 3007 & 69.40 & 46.67\% & $29.53\pm29.00$\\
P3 & \bf{3002} & 74.62 & 43.41\% & $38.06\pm27.85$\\
P4 & 3004 & 61.51 & 52.66\% & $44.51\pm9.46$\\
P5 & 3004 & 62.85 & 51.54\% & $65.98\pm13.04$\\
A$^{\dagger}$ & N/A & N/A & N/A & $106.73\pm63.01$\\
B$^{\ddagger}$ & 3024 & 58.48 & 55.39\% & $21.47\pm7.57$\\
C$^{\ddagger}$ & 3019 & 64.10 & 50.53\% & $21.47\pm7.52$\\
D$^{\mathsection}$ & 3388 & 74.04 & 43.75\% & $23.46\pm9.14$\\
E$^*$ & 3846 & 82.82 & 39.11\% & $27.90\pm14.21$\\
\bottomrule
\end{tabular}
\begin{tablenotes}
\small
\item[$\dagger$] Infeasible until simulation timeouts
\item[$\ddagger$] Infeasible at 0 - 5 s
\item[$\mathsection$] Infeasible at 0 - 5 s, 755 - 1125 s; violates $\overline{T}_{\mathrm{core}}$
\item[$*$] Infeasible at 0 - 5 s, 600 - 1430 s; violates $\overline{T}_{\mathrm{core}}$
\end{tablenotes}
\end{threeparttable}
\centering
\end{table}

\begin{figure*}
\subfloat[Result by P1 with $T_{\mathrm{amb}} = -25$ $^{\circ}\text{C}$.]
{\label{fig:temp_higher_bound:P1} \includegraphics[width=0.32\linewidth]{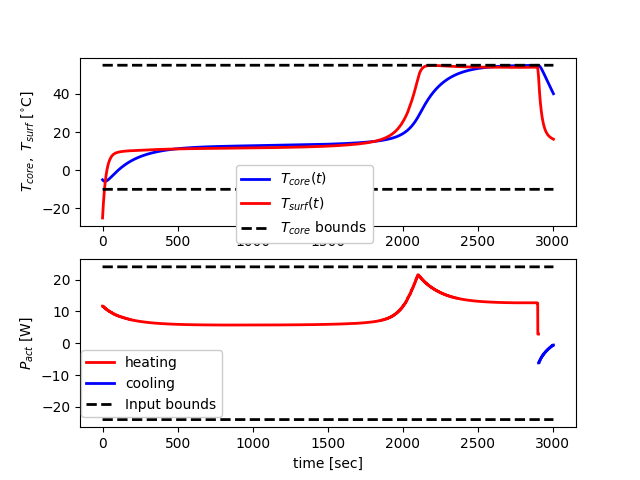}}
\hfill
\subfloat[Result by P3 with $T_{\mathrm{amb}} = -25$ $^{\circ}\text{C}$.]
{\label{fig:temp_higher_bound:P3} \includegraphics[width=0.295\linewidth]{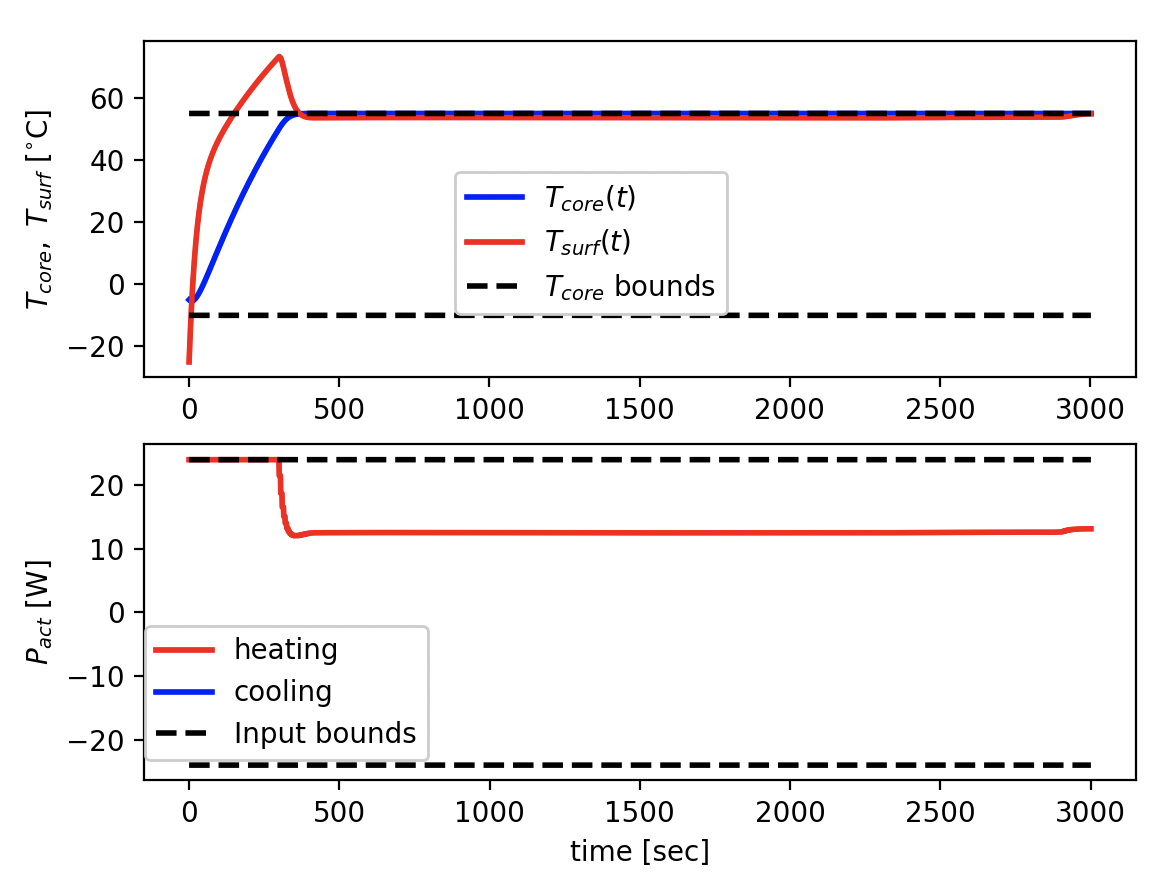}}
\hfill
\subfloat[Result by P5 with $T_{\mathrm{amb}} = -25$ $^{\circ}\text{C}$.]
{\label{fig:temp_higher_bound:P5} \includegraphics[width=0.32\linewidth]{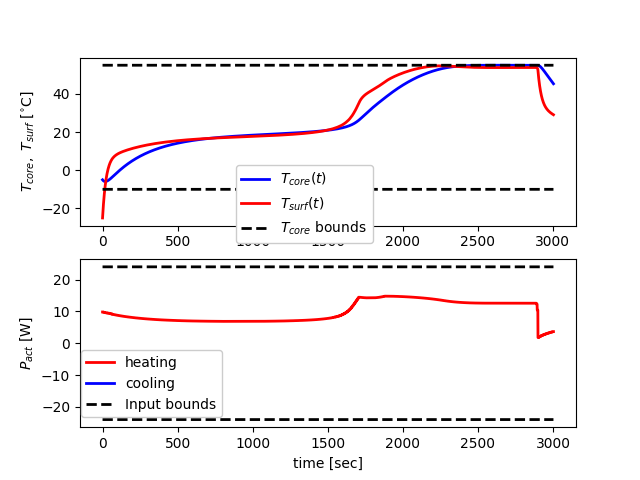}}
\caption{Battery core and surface temperature and active thermal power by strategies P1, P3, and P5 with larger thermal power bounds in the low ambient temperature. Note that there is no constraint on $T_{\mathrm{surf}}$.} \label{fig:temp_higher_bound}
\end{figure*}

As for the effect of $T_{\mathrm{core}}$ on charging speed, P3 outperforms the other strategies in terms of charging time.
Meanwhile, the other proposed strategies yield slightly longer charging times.
Nevertheless, P1 can achieve a good balance across all indices.
Fig.~\ref{fig:temp_higher_bound:P3} illustrates the reason why P3 outperforms the others on charging time, where P3 tracks $T_{\mathrm{core,r}}=55$ $^{\circ}\text{C}$ to maintain a high $T_{\mathrm{core}}$ for a longer duration than P1 and P5.
The reason can be revealed by the proposed thermal-NDC model, where a higher $T_{\mathrm{core}}$ reduces the internal resistance $R_{\mathrm{b,T}}(t)$ as \eqref{eq:diffusion_resis} describes, and further affects the capacitor voltage dynamics \eqref{eq:NDC_model:two_v}.
Since $\mathrm{SoC}(t)$ is defined as a linear combination of the capacitor voltages $V_{\mathrm{b}}(t)$ and $V_{\mathrm{s}}(t)$ in \eqref{eq:soc_define}, the time derivative of SoC is directly affected by the capacitor voltage dynamics \eqref{eq:NDC_model:two_v}. This explanation is consistent with the observation from numerical experiments, as well as the conclusions from the literature \cite{wang2016lithium,yang2018fast}.
However, heating the battery at the beginning as P3 does is not energy-efficient. Fig.~\ref{fig:temp_higher_bound:P1} and Fig.~\ref{fig:temp_higher_bound:P5} indicate that after the battery is charged for about 2200 s and 1700 s, P1 and P5 begin to heat the battery because they generate more heat by \eqref{eq:heat_generation_rate} with a higher $V$.

Comparing the results from Table~\ref{table:higher_power_low_temp} and Fig.~\ref{fig:temp_higher_bound}, P1 only increases the heating power after $\sim$2100 seconds, where the SoC is $\sim$65\%.
From the comparison of the MPC horizon $N$ in this subsection, the horizon of P (or P1) is long enough such that the MPC can obtain a solution that is at least close to the optimal one. Thus, explicitly regulating $T_{\mathrm{core}}$ as P2 and P3 do is not necessary.
From the perspective of balancing the charging time and energy consumption, P1 performs better than the other strategies.

\section{EKF-Based Output-Feedback Model Predictive Control} \label{sec:ekf_mpc}

The previous section studies the MPC of battery charging based on the assumption that the states are fully measurable.
This assumption might not hold in real-world applications. Hence, this section proposes an output-feedback MPC based on Extended Kalman Filter (EKF) \cite[Chapter~8.2]{anderson2012optimal}. The EKF takes in system output measurements and generates state estimates, which are given to a revised MPC controller to design control. This section presents a way to rewrite the proposed thermal-NDC model for the ease of EKF, proposes an EKF-based output-feedback MPC, and demonstrates the effectiveness of this algorithm under extreme ambient temperatures by numerical simulations.

\subsection{System Reformulation \& Proposed Algorithm}

The output $V$ is a function of both states and inputs, as shown in \eqref{eq:NDC_model:one_terminal_v}, thus EKF is not applicable for this output form. This subsection first rewrites the original system dynamics \eqref{eq:system_dynamics} similarly as \cite{tian2020real} such that its output is a function of states.
To reformulate the system dynamics, new state, control, and output for the continuous-time system dynamics are defined as follows,
\begin{equation} \label{eq:ekf_state_define}
\begin{split}
&\boldsymbol{x}(t) \triangleq \text{col}\{ V_{\mathrm{b}}, V_{\mathrm{s}}, T_{\mathrm{core}}, T_{\mathrm{surf}}, I \} \in \mathbb{R}^5, \\
&\boldsymbol{u}(t) \triangleq \text{col}\{ u_1, P_{\mathrm{act}} \}  \in \mathbb{R}^2, \\
&\boldsymbol{y}(t) \triangleq \text{col}\{ T_{\mathrm{surf}}, V, I \} \in \mathbb{R}^3,
\end{split}
\end{equation}
where $u_1 \in \mathbb{R}$ is a new control input such that the charging current has the following fictitious dynamics:
\begin{equation} \label{eq:current_dyn}
\dot{I}(t) = u_1(t).
\end{equation}
Note that $\boldsymbol{y}(t)$ is measurable in practice.

Combining the previous dynamics \eqref{eq:NDC_model} - \eqref{eq:heat_active} with the current dynamics \eqref{eq:current_dyn}, one obtains the following dynamics in continuous-time:
\begin{subequations} \label{eq:control_affine_dyn}
\begin{align}
\dot{\boldsymbol{x}} &= \hat{\boldsymbol{f}}_{\mathrm{c}}(\boldsymbol{x},\boldsymbol{u}), \\
\boldsymbol{y} &= \boldsymbol{h}(\boldsymbol{x}),
\end{align}
\end{subequations}
where $\hat{\boldsymbol{f}}_{\mathrm{c}}: \mathbb{R}^5 \times \mathbb{R}^2 \mapsto \mathbb{R}^5$ and $\boldsymbol{h}: \mathbb{R}^5 \mapsto \mathbb{R}^3$ is the new output mapping.
Thus, the discrete-time system dynamics for EKF are given by:
\begin{equation} \label{eq:system_dynamics_kalman}
\begin{split}
\boldsymbol{x}_{i+1}& = \boldsymbol{f}_{\mathrm{d}}(\boldsymbol{x}_{i}, \boldsymbol{u}_i) + \boldsymbol{w}_i \coloneqq \boldsymbol{x}_i + \Delta_{\mathrm{s}}\hat{\boldsymbol{f}}_{\mathrm{c}}(\boldsymbol{x}_i,\boldsymbol{u}_i) + \boldsymbol{w}_i, \\
\boldsymbol{y}_i &= \boldsymbol{h}(\boldsymbol{x}_i) + \boldsymbol{v}_i, \\
\end{split}
\end{equation}
where $\boldsymbol{f}_{\mathrm{d}}: \mathbb{R}^5 \times \mathbb{R}^2 \mapsto \mathbb{R}^5$; $\boldsymbol{w}_i$ and $\boldsymbol{v}_i$ are the process disturbance and observation noise, respectively; $\boldsymbol{w}_i$ and $\boldsymbol{v}_i$ are both assumed to be zero mean multivariate Gaussian noises with covariance matrices $\boldsymbol{Q}_i \in \mathbb{R}^{5 \times 5}$ and $\boldsymbol{R}_i\in \mathbb{R}^{3 \times 3}$, respectively.

The EKF for the battery system \eqref{eq:system_dynamics_kalman} is summarized in Algorithm \ref{alg:ekf}.
Denote $\boldsymbol{F}_i \triangleq \frac{\partial \boldsymbol{f}_{\mathrm{d}}(\boldsymbol{x},\boldsymbol{u})}{\partial \boldsymbol{x}} |_{\hat{\boldsymbol{x}}_{i-1|i-1}, \boldsymbol{u}_{i-1}}$ and $\boldsymbol{H}_i \triangleq \frac{\partial \boldsymbol{h}(\boldsymbol{x})}{\partial \boldsymbol{x}} |_{\hat{\boldsymbol{x}}_{i-1|i-1}}.$
At each time instance $t_i$, the system output $\boldsymbol{y}_i$ is observed. Then with the previous state estimate $\hat{\boldsymbol{x}}_{i-1|i-1}$, covariance estimate $\boldsymbol{P}_{i-1|i-1}$, and control input $\boldsymbol{u}_{i-1}$, the EKF updates the present estimates $\hat{\boldsymbol{x}}_{i|i}$ and $\boldsymbol{P}_{i|i}$. Note that the estimated SoC at time $t_i$ is given by its definition \eqref{eq:soc_define}, i.e., $\hat{\mathrm{SoC}}(t_i) = (C_{\mathrm{b}} \hat{V}_{\mathrm{b}}(t_i) + C_{\mathrm{s}} \hat{V}_{\mathrm{s}}(t_i))/(C_{\mathrm{b}}\overline{V}_{\mathrm{b}} + C_{\mathrm{s}}\overline{V}_{\mathrm{s}})$.
\begin{algorithm}
\caption{Extended Kalman Filter (EKF)}\label{alg:ekf}
\DontPrintSemicolon
\KwIn{system \eqref{eq:system_dynamics_kalman}, $\boldsymbol{Q}_i$, $\boldsymbol{R}_i$}
\SetKwFunction{estimate}{estimate}
\SetKwProg{Fn}{def}{:}{}
\Fn{\estimate{$\hat{\boldsymbol{x}}_{i-1|i-1}, \boldsymbol{P}_{i-1|i-1}, \boldsymbol{u}_{i-1}, \boldsymbol{y}_i$}}{
$\hat{\boldsymbol{x}}_{i|i-1} \gets \boldsymbol{f}_{\mathrm{d}}(\hat{\boldsymbol{x}}_{i-1|i-1}, \boldsymbol{u}_{i-1})$\;
$\boldsymbol{P}_{i|i-1} = \boldsymbol{F}_i \boldsymbol{P}_{i-1|i-1} \boldsymbol{F}_i^{\top} + \boldsymbol{Q}_i$\;
$\tilde{\boldsymbol{y}}_i \gets \boldsymbol{y}_i - \boldsymbol{h}(\hat{\boldsymbol{x}}_{i|i-1})$\;
$\boldsymbol{S}_i \gets \boldsymbol{H}_i \boldsymbol{P}_{i|i-1} \boldsymbol{H}_k^{\top} + \boldsymbol{R}_i$\;
$\boldsymbol{K}_i \gets \boldsymbol{P}_{i|i-1}\boldsymbol{H}_i^{\top}\boldsymbol{S}_i^{-1}$\;
$\hat{\boldsymbol{x}}_{i|i} = \hat{\boldsymbol{x}}_{i|i-1} + \boldsymbol{K}_i \tilde{\boldsymbol{y}}_i$\;
$\boldsymbol{P}_{i|i} = (\boldsymbol{I}_{5} - \boldsymbol{K}_i\boldsymbol{H}_i)\boldsymbol{P}_{i|i-1}$\;
\KwRet $\hat{\boldsymbol{x}}_{i|i},\boldsymbol{P}_{i|i}$\;
}
\end{algorithm}

Finally, at each time $t_k$, the optimal control can be determined by an EKF-based output-feedback MPC given the present EKF estimate $\hat{\boldsymbol{x}}_{k|k}$. The MPC formulation at $t_k$ is the same as \eqref{eq:oc_mpc_1}, except the initial state $\boldsymbol{x}_{k}$ replaced by $\hat{\boldsymbol{x}}_{k|k}$, and the system dynamics within the MPC replaced by
\begin{equation}
\boldsymbol{x}_{k+j+1|k} = \boldsymbol{x}_{k+j|k} + \Delta_{\mathrm{p}} \hat{\boldsymbol{f}}_{\mathrm{c}}(\boldsymbol{x}_{k+j|k}, \boldsymbol{u}_{k+j|k}).
\end{equation}
The entire EKF-based output-feedback MPC is summarized in Algorithm \ref{alg:mpc_output_ekf}. To compensate for the uncertainty and measurement noise, the Li-ion concentration gradient constraint \eqref{constraint:V_s_minus_V_b_old} is revised conservatively as
\begin{equation} \label{constraint:V_s_minus_V_b_old:revised}
V_{\mathrm{s}}(t) - V_{\mathrm{b}}(t) \leq \beta_1 (\mathrm{SoC}(t) + 5\%) + \beta_2, \ \forall t,
\end{equation}
for a safety margin of 5\%.
Section~\ref{subsec:sim_ekf} discusses how this constraint affects the output-feedback MPC and the charging performance.




\begin{algorithm}
\caption{EKF-Based Output-Feedback MPC}\label{alg:mpc_output_ekf}
\DontPrintSemicolon
\KwIn{$k = -1$, $i=0$, $t_0=0$, $\Delta_{\mathrm{p}}$, $\Delta_{\mathrm{s}}$, $I_0=0$}
initialize $\hat{\boldsymbol{x}}_{i-1|i-1}$ and $\boldsymbol{P}_{i-1|i-1}$\;
\While {$\hat{\mathrm{SoC}}_i < \mathrm{SoC}_{\mathrm{r}}$} {
observe $\boldsymbol{y}_i$ and $T_{\mathrm{amb},i}$ at time $t_i$\;
$\hat{\boldsymbol{x}}_{i|i},\boldsymbol{P}_{i|i} \gets$ \estimate{$\hat{\boldsymbol{x}}_{i-1|i-1}, \boldsymbol{P}_{i-1|i-1}, \boldsymbol{u}_{i-1}, \boldsymbol{y}_i$}\;
obtain $\hat{\mathrm{SoC}}_i$ given $\hat{\boldsymbol{x}}_{i|i}$\;

\uIf{$t_i \ \% \ \Delta_{\mathrm{p}} == 0$}
{
$k \gets k + 1$\;
$\boldsymbol{u}^*_{0:N-1|k} \gets$  solve $\eqref{eq:oc_mpc_1}$ given $\hat{\boldsymbol{x}}_{k|k}$ at time $t_k$ and $T_{\mathrm{amb},k+j|k} \equiv T_{\mathrm{amb},k}, \  \forall j \in \llbracket 0,N \rrbracket$\;
$t_k \gets t_k + \Delta_{\mathrm{p}}$
\;}
\Else{
perform $\boldsymbol{u}_i$ by interpolating $\boldsymbol{u}^*_{0:N-1|k}$ with present time $t_i$ and zero-order hold\;
}
$t_{i} \gets t_i + \Delta_{\mathrm{s}}$, $i \gets i+1$\;
}
\end{algorithm}

\subsection{Numerical Simulations} \label{subsec:sim_ekf}

This subsection presents numerical simulations on EKF-based output-feedback MPC for battery charging under extreme ambient temperatures.
In this subsection, parameters are the same as the strategy P1 in Section \ref{subsec:basic_case_study}, except for explicit annotations.
The discrete-time system dynamics in the simulation is updated in every time interval $\Delta_{\mathrm{s}}$ and given by
\begin{equation}
\boldsymbol{x}_{i+1} =\boldsymbol{x}_{i} + \Delta_{\mathrm{s}} \hat{\boldsymbol{f}}_{\mathrm{c}}(\boldsymbol{x}_i, \boldsymbol{u}_i).
\end{equation}

The covariance matrix $\boldsymbol{Q}_i$ for the process noise is set based on the relative modeling error or process uncertainty.
For $V_{\mathrm{b}}$ and $V_{\mathrm{s}}$, the dominant source of error is the capacitor $C_{\mathrm{b}}$, as described in \eqref{eq:NDC_model:one_terminal_v}.
To account for a relative error of $\pm$ 7.5\% with a probability of 99.7\%, the variance for $V_{\mathrm{b}}$ in discrete-time should be $(2.5\% \Delta_{\mathrm{s}}\frac{-1}{C_{\mathrm{b}} R_{\mathrm{b}}})^2 = 1.73 \times 10^{-8}$. The variance for $C_{\mathrm{s}}$ is also $1.73 \times 10^{-8}$ for conservatism.
For $T_{\mathrm{core}}$ and $T_{\mathrm{surf}}$, according to \eqref{eq:lumped_thermal_model} and the measurability of $T_{\mathrm{surf}}$, the dominant source of error is $T_{\mathrm{core}}$. To cover a relative error of $\pm$ 7.5\% with a probability of 99.7\%, the variance for $T_{\mathrm{core}}$ should be $(2.5\% \Delta_{\mathrm{s}}\frac{-1}{R_{\mathrm{core}}C_{\mathrm{core}}})^2 = 2.44 \times 10^{-8}$.
For $T_{\mathrm{surf}}$, to cover a relative error of $\pm$ 0.3\%, the variance should be $(0.1\% \Delta_{\mathrm{s}} (\frac{-1}{R_{\mathrm{surf}}C_{\mathrm{surf}}} + \frac{-1}{R_{\mathrm{core}}C_{\mathrm{surf}}}))^2 = 1.54\times 10^{-9}$.
Since the current dynamics \eqref{eq:current_dyn} is manually constructed to enable EKF, there is no modeling error or uncertainty on $I$, thus the variance for $I$ is 0.
Therefore, the process covariance matrix is $\boldsymbol{Q}_i = \text{diag}(1.73 \times 10^{-8}, \ 1.73 \times 10^{-8}, \ 2.44 \times 10^{-8}, \ 1.54\times 10^{-9}, \ 0)$.

The Gaussian noise covariance matrix $\boldsymbol{R}_i = \text{diag}(10^{-3}, \ 10^{-5}, \ 10^{-12})$, where each component's corresponding standard deviation is 0.032 $^{\circ}\text{C}$, 0.0032 V, and $1\times 10^{-6}$ A, respectively. Thus, the error range with a probability of 99.7\% for each component is about $\pm 0.1$ $^{\circ}\text{C}$, $\pm$0.01 V, and $\pm 3\times 10^{-6}$ A.
Note that the covariance of $I$ is small because the charging current $I$ is actually a control input. This section reformulates the previous dynamics \eqref{eq:system_dynamics} to enable an EKF. Eventually, a charging current will be applied. Thus, the measurement noise on $I$ is assumed to be small.
The measurement $\boldsymbol{y}_i$ at each time instance $t_i$ is given by the true output corresponding to $\boldsymbol{x}_i$ with Gaussian noise sampled with $\boldsymbol{R}_i$.
The initial covariance estimate $\boldsymbol{P}_{i-1|i-1} = \text{diag}(0.5,\  0.5, \ 0.5, \ 0.01, \ 0.01)$, where the small covariance elements are given because the corresponding states are measurable.

\begin{figure}[ht]
\centering
\includegraphics[width=0.35\textwidth]{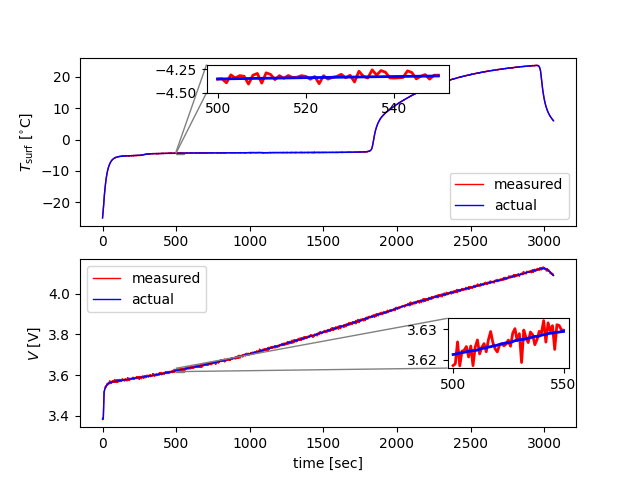}
\caption{Actual and measured system outputs with $T_{\mathrm{amb}} = -25$ $^{\circ}\text{C}$.}
\label{fig:measure_ekf_low}
\end{figure}

The trajectory of measured and actual outputs with $T_{\mathrm{amb}} = -25$ $^{\circ}\text{C}$ is illustrated by Fig~\ref{fig:measure_ekf_low}. The measurements with different $T_{\mathrm{amb}}$ are similar to Fig~\ref{fig:measure_ekf_low}, thereby omitted.
The other parameters are summarized below.
\begin{enumerate}
\item Mild Ambient Temperature: $T_{\mathrm{amb}} = 25$ $^{\circ}\text{C}$, $\boldsymbol{x}_0 = \matt{0.1 \text{ V} & 0.1 \text{ V} & 25\,^{\circ}\text{C} & 25\,^{\circ}\text{C} & 0 \text{ A}}^{\top}$;
\item High Ambient Temperature: $T_{\mathrm{amb}} = 70$ $^{\circ}\text{C}$, $\boldsymbol{x}_0 = \matt{0.1 \text{ V} & 0.1 \text{ V} & 50\,^{\circ}\text{C} & 70\,^{\circ}\text{C} & 0 \text{ A}}^{\top}$;
\item Low Ambient Temperature: $T_{\mathrm{amb}} = -25$ $^{\circ}\text{C}$, $\boldsymbol{x}_0 = \matt{0.1 \text{ V} & 0.1 \text{ V} & -5\,^{\circ}\text{C} & -25\,^{\circ}\text{C} & 0 \text{ A}}^{\top}$.
\end{enumerate}


\begin{figure}[ht]
\centering
\includegraphics[width=0.49\textwidth]{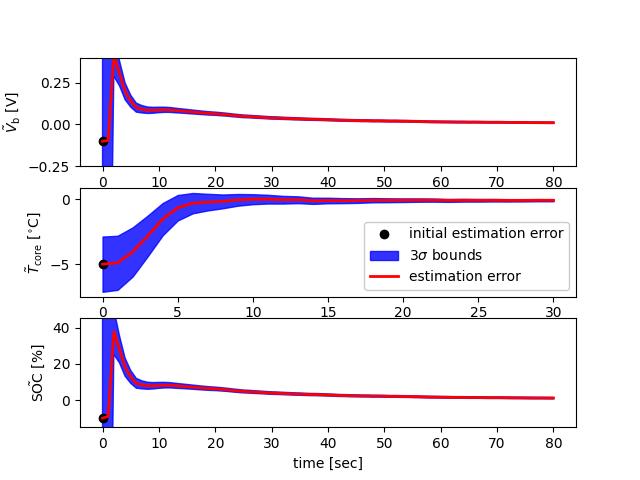}
\caption{The trajectory segment of the estimation errors on unmeasurable states $V_{\mathrm{b}}$, $T_{\mathrm{core}}$ and $\mathrm{SoC}$ with $T_{\mathrm{amb}} = 25$ $^{\circ}\text{C}$.}
\label{fig:estimate_ekf_mild}
\end{figure}

\begin{figure}[ht]
\centering
\includegraphics[width=0.49\textwidth]{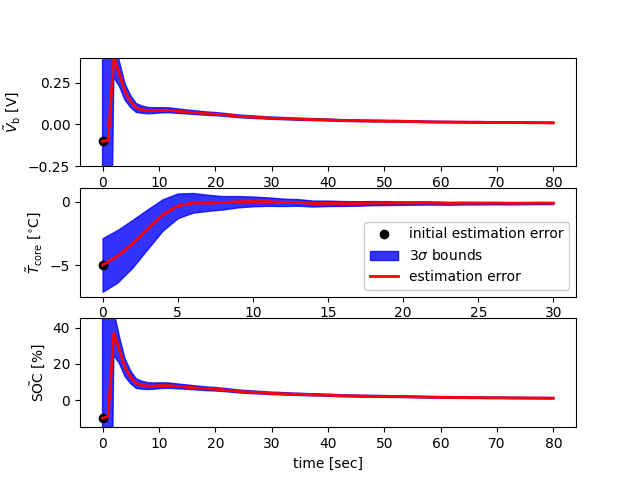}
\caption{The trajectory segment of the estimation errors on unmeasurable states $V_{\mathrm{b}}$, $T_{\mathrm{core}}$ and $\mathrm{SoC}$ with $T_{\mathrm{amb}} = 70$ $^{\circ}\text{C}$.}
\label{fig:estimate_ekf_high}
\end{figure}

\begin{figure}[ht]
\centering
\includegraphics[width=0.49\textwidth]{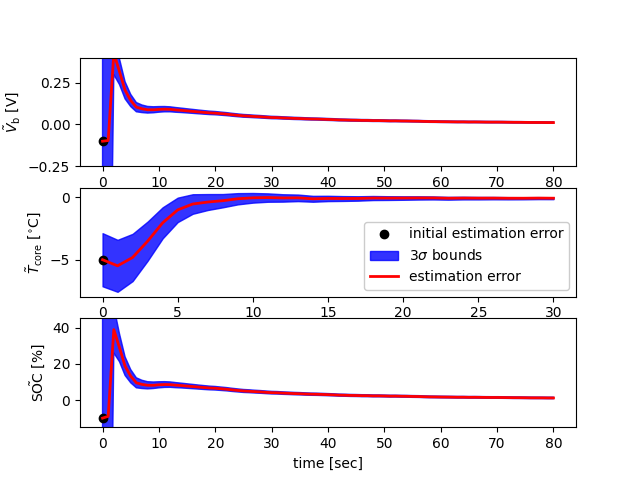}
\caption{The trajectory segment of the estimation errors on unmeasurable states $V_{\mathrm{b}}$, $T_{\mathrm{core}}$ and $\mathrm{SoC}$ with $T_{\mathrm{amb}} = -25$ $^{\circ}\text{C}$.}
\label{fig:estimate_ekf_low}
\end{figure}



The initial estimates on the unmeasurable states $V_{\mathrm{b}}$ and $T_{\mathrm{core}}$ are randomly initialized within $[V_{\mathrm{b}}(0)-0.1 \text{ V}, V_{\mathrm{b}}(0)+0.1 \text{ V}]$ and $[T_{\mathrm{core}}(0)-5 \ ^{\circ}\text{C}, T_{\mathrm{core}}(0)+5 \ ^{\circ}\text{C}]$, respectively. Note that the initial estimate error on $V_{\mathrm{b}}$ is relatively large since $V_{\mathrm{b}}$ varies in $[0 \text{ V}, 1 \text{ V}]$. The initial estimates of the measurable states are given by the initial measurements.
For each $T_{\mathrm{amb}}$, the proposed EKF-based output-feedback MPC runs 20 trials with random initial estimates to verify its effectiveness.

The estimation errors for the unmeasurable states $V_{\mathrm{b}}$, $T_{\mathrm{core}}$ and $\mathrm{SoC}$ with different $T_{\mathrm{amb}}$ are illustrated in Fig.~\ref{fig:estimate_ekf_mild} - Fig.~\ref{fig:estimate_ekf_low}, where black dots indicate the initial state estimate; blue buffers indicate the estimation error plus and minus 3 times of standard deviation $\sigma$ given the estimate covariance $\boldsymbol{P}_{i|i}$.
A small 3$\sigma$ error bound indicates that a 99.7\% confidence interval of the state estimate is compact around the true state value.
According to Fig.~\ref{fig:estimate_ekf_mild} - Fig.~\ref{fig:estimate_ekf_low}, both the state estimates and the $3\sigma$ error bound converge within about 80 s.
Table~\ref{table:ekf_estimation_error} summarizes the EKF estimation error for unmeasurable states, which validates the EKF's effectiveness.

\begin{table}
\centering
\begin{threeparttable}
\caption{EKF Estimation Error} \label{table:ekf_estimation_error}
\begin{tabular}{c c c c c}
\toprule
State$^\dagger$ & $T_{\mathrm{amb}}$ & mean $\pm$ std & quartiles$^\ddagger$ \\
\midrule
$V_{\mathrm{b}}$ & 25 $^{\circ}$C & $0.0012\pm0.0064$ & 0.0003, 0.0007, 0.0011 \\
$V_{\mathrm{s}}$ & 25 $^{\circ}$C & $0.0008\pm0.0018$ & 0.0003, 0.0006, 0.0009 \\
$T_{\mathrm{core}}$ & 25 $^{\circ}$C & $0.0172\pm0.1051$ & 0.0013, 0.0030, 0.0092 \\
$\mathrm{SoC}$ & 25 $^{\circ}$C & $0.11\%\pm0.59\%$ & 0.03\%, 0.06\%, 0.11\% \\
\midrule
$V_{\mathrm{b}}$ & 70 $^{\circ}$C & $0.0012\pm0.0065$ & 0.0003, 0.0007, 0.0011 \\
$V_{\mathrm{s}}$ & 70 $^{\circ}$C & $0.0008\pm0.0018$ & 0.0003, 0.0006, 0.0009 \\
$T_{\mathrm{core}}$ & 70 $^{\circ}$C & $0.0092\pm0.1070$ & 0.0011, 0.0023, 0.0047 \\
$\mathrm{SoC}$ & 70 $^{\circ}$C & $0.11\%\pm0.60\%$ & 0.03\%, 0.06\%, 0.11\% \\
\midrule
$V_{\mathrm{b}}$ & -25 $^{\circ}$C & $0.0012\pm0.0067$ & 0.0003, 0.0006, 0.0011 \\
$V_{\mathrm{s}}$ & -25 $^{\circ}$C & $0.0008\pm0.0018$ & 0.0003, 0.0005, 0.0009 \\
$T_{\mathrm{core}}$ & -25 $^{\circ}$C & $0.0174\pm0.1087$ & 0.0018, 0.0056, 0.0186 \\
$\mathrm{SoC}$ & -25 $^{\circ}$C & $0.11\%\pm0.62\%$ & 0.03\%, 0.06\%, 0.11\% \\
\bottomrule
\end{tabular}
\begin{tablenotes}
\small
\item[$\dagger$] error unit for $V_{\mathrm{b}}$, $V_{\mathrm{s}}$, $T_{\mathrm{core}}$, and $\mathrm{SoC}$ are V, V, K, and \%, respectively
\item[$\ddagger$] 25th, 50th, 75th percentile
\end{tablenotes}
\end{threeparttable}
\centering
\end{table}

\begin{figure}[ht]
\centering
\includegraphics[width=0.49\textwidth]{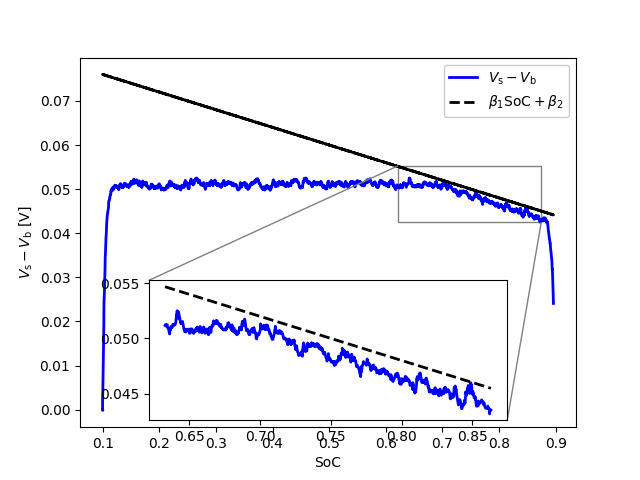}
\caption{The trajectory of  $V_{\mathrm{s}}(t) - V_{\mathrm{b}}(t)$ vs $\beta_1 \mathrm{SoC}(t) + \beta_2$ with $T_{\mathrm{amb}} = 70$ $^{\circ}\text{C}$.}
\label{fig:zeta_high}
\end{figure}

The charging performance for all cases is summarized in Table~\ref{table:ekf_mpc_result}.
Comparing with the performance of P1 in Table~\ref{table:basic_mild} - Table~\ref{table:basic_low}, both the computational time and the charging time from Table~\ref{table:ekf_mpc_result} are longer than those of the state-feedback MPC.
One contributing factor is the conservative revision of the Li-ion concentration gradient constraint \eqref{constraint:V_s_minus_V_b_old:revised}, which includes a safety margin to account for estimation error and measurement noise.
To illustrate this, consider the trajectory of  $V_{\mathrm{s}}(t) - V_{\mathrm{b}}(t)$ vs $\beta_1 \mathrm{SoC}(t) + \beta_2$ with $T_{\mathrm{amb}} = 70$ $^{\circ}$C, as shown in Fig.~\ref{fig:zeta_high}. In Fig.~\ref{fig:extra}, the same trajectory of $V_{\mathrm{s}}(t) - V_{\mathrm{b}}(t)$ without any noise or uncertainty  approaches the boundary of $\beta_1 \mathrm{SoC}(t) + \beta_2$ as SoC exceeds about 70\%.
However, when using output-feedback MPC with noise and uncertainty, the actual $V_{\mathrm{s}}(t) - V_{\mathrm{b}}(t)$ could exceed this boundary. Therefore, the 5\% safety margin in \eqref{constraint:V_s_minus_V_b_old:revised} prevents constraint violation, but it also slows down the charging process due to the reduced concentration gradient.

On the other hand, the charging success rate is 100\% for all the trials.
As for constraint violations, for three $T_{\mathrm{amb}}$, the average time duration percentage that violates constraints is 0.0\%$\pm$0.0\%, 0.0\%$\pm$0.0\%, and 0.0033\%$\pm$0.0144\%, respectively. The maximum constraint violation is less than 0.1\% of its upper bound, which might be caused by the measurement or numerical noises.
The numerical experiments above validate the effectiveness of the proposed EKF-based output-feedback MPC under extreme ambient temperatures.

In this subsection, the proposed output-feedback MPC directly cold-starts with a relatively large estimation error.
One practical approach is to initially run EKF independently for a period, allowing it to reduce the initial estimation error. Subsequently, the MPC algorithm can be initiated for charging.
Details are referred to in the literature related to the separation principle \cite{wonham1968separation,atassi1999separation,georgiou2013separation}.


\begin{table}
\centering
\begin{threeparttable}
\caption{Battery Charging Result with EKF-Based Output-Feedback MPC} \label{table:ekf_mpc_result}
\begin{tabular}{c c c c c c}
\toprule
$T_{\mathrm{chg}}$ [s] & Energy [kJ] & Efficiency [\%] & $T_{\mathrm{comp}}$ [ms] \\
\midrule
3019.55$\pm$21.75 & 39.64$\pm$0.27 & 81.81$\pm$0.08 & $30.41\pm5.37$ \\
\midrule
3017.30$\pm$20.90 & 44.43$\pm$0.31 & 72.98$\pm$0.03 & $27.94\pm4.53$ \\
\midrule
3041.30$\pm$20.31 & 48.30$\pm$0.32 & 67.18$\pm$0.06 & $33.09\pm7.02$ \\
\bottomrule
\end{tabular}
\begin{tablenotes}
\small
\item $T_{\mathrm{amb}} = 25\  ^{\circ}\text{C}, \ 70\  ^{\circ}\text{C}, \ -25\  ^{\circ}\text{C}$ from top to bottom
\end{tablenotes}
\end{threeparttable}
\centering
\end{table}


\section{Conclusion and Discussion} \label{sec:conclusion}

This paper explores integrated control strategies for optimal fast charging and active thermal management of LiBs under extreme ambient temperatures, aiming to balance charging speed and battery health.
First, a control-oriented thermal-NDC model is introduced to describe the electrical and thermal dynamics of LiBs, including the impact from both an active thermal source and ambient temperature.
Second, a state-feedback MPC algorithm is developed to perform optimal fast charging and active thermal management concurrently.
Third, numerical experiments are conducted to validate the proposed state-feedback MPC algorithm under extreme ambient temperatures. These experiments demonstrate that heating the battery can enhance fast charging, which is consistent with the literature.
By explicitly incorporating a higher desired battery core temperature into the MPC's objective function, the algorithm achieves faster charging speed.
However, with a suitable prediction horizon, an objective function without explicit heating can achieve a balanced performance across various indices. Finally, the proposed thermal-NDC model is rewritten properly to enable an EKF design for battery state estimation. An EKF-based output-feedback MPC algorithm is then proposed and validated through numerical experiments under extreme ambient temperatures.

The limitations of this paper can be summarized from three key perspectives.
First, the proposed thermal-NDC model is validated for charging and discharging currents up to 1.5C. This limitation arises because the model is based on experimental data collected at a maximum current of 1.5C, limiting its applicability. As highlighted by \cite{yang2021challenges,biju2023battx}, applications such as eVTOL aircraft and EVs often require high C-rate charging, which generates significant heat and raises concerns about system safety.
However, the methodology presented in this paper could be readily adapted to higher C-rate charging if a more sophisticated model, suitable for high C-rates, is used.
Therefore, a future work direction is to develop a battery model and corresponding control strategy capable of handling higher currents, such as up to 4-6C.


Second, the lumped thermal model used in this paper may lose some fidelity, as higher-fidelity thermal dynamics are described by PDEs \cite{zeng2021review}.
Nevertheless, lumped thermal models, governed by ODEs, are frequently used in the literature \cite{lin2014lumped, perez2017optimal,zou2017electrothermal,dong2024optimal,goshtasbi2024evtol,hamednia2023optimal,tu2024system} due to the computational feasibility for real-time model-based control.
Some studies utilize two-point or multiple-point lumped models to capture the thermal dynamics of battery packs \cite{goshtasbi2024evtol,hamednia2023optimal}.
Although the literature \cite{zeng2021review} and experimental data \cite{lin2014lumped} suggest that lumped thermal models remain valid for control purposes, this paper acknowledges the modeling mismatch between lumped thermal models and actual thermal dynamics.
This discrepancy highlights the need to design robust control strategies to address modeling errors and process uncertainties in future work.

Third, the dynamics modeling of the heating/cooling system is omitted in this paper, simplifying the active thermal power as \eqref{eq:heat_active}.
Although modeling the heating/cooling system with actuator dynamics (e.g., heat pump, condenser, evaporator, fan, etc.) provides valuable insights into system behavior and benefits model-based thermal management \cite{park2020computationally,park2021model,hajidavalloo2023nmpc}, the additional complexity would detract from the primary focus of this paper.
The advantage of using \eqref{eq:heat_active} is that it simplifies the modeling and control of the battery cell and thermal system by hierarchically separating them. This approach allows the proposed algorithm to plan and determine the desired active thermal power for the battery cell. Subsequently, a separate control strategy can regulate the thermal system's actuators to achieve this desired thermal power.
Another benefit of using \eqref{eq:heat_active} is modularization. As noted in \cite{goshtasbi2024evtol}, the lumped model for a battery cell can be extended as a multi-point lumped model for a battery pack, where one node of the battery cells receives active thermal power through cold plates. The above justifications make the lumped model for thermal dynamics a reasonable choice for battery cell electrothermal dynamics and model-based control strategies.
Regarding the typical range for the value of $\eta_{\mathrm{act}}$ in \eqref{eq:heat_active}, it is difficult to specify one because this efficiency depends on various factors, such as the configuration of the heating/cooling system, the topology of the battery pack, actuator types, etc. Furthermore, as noted in \cite{park2020computationally,park2021model,hajidavalloo2023nmpc}, this efficiency is typically a state variable of the thermal system, influenced by fan speed, pump speed, coolant temperature, and other factors.
Therefore, if a constant value for $\eta_{\mathrm{act}}$ cannot be predetermined, $\dot{Q}_{\mathrm{act}}(t)$ can be made as a decision variable. The proposed strategy can then determine a trajectory for the desired net active thermal power. A separate strategy from the heating/cooling system can then be employed to track this desired power trajectory as a reference trajectory.
Nevertheless, this limitation underscores the need for future work to model the thermal dynamics and the heating/cooling dynamics (including efficiency) for a battery pack.

\bibliographystyle{IEEEtran}
\bibliography{Reference}

\begin{thebibliography}{10}
\providecommand{\url}[1]{#1}
\csname url@samestyle\endcsname
\providecommand{\newblock}{\relax}
\providecommand{\bibinfo}[2]{#2}
\providecommand{\BIBentrySTDinterwordspacing}{\spaceskip=0pt\relax}
\providecommand{\BIBentryALTinterwordstretchfactor}{4}
\providecommand{\BIBentryALTinterwordspacing}{\spaceskip=\fontdimen2\font plus
\BIBentryALTinterwordstretchfactor\fontdimen3\font minus
  \fontdimen4\font\relax}
\providecommand{\BIBforeignlanguage}[2]{{%
\expandafter\ifx\csname l@#1\endcsname\relax
\typeout{** WARNING: IEEEtran.bst: No hyphenation pattern has been}%
\typeout{** loaded for the language `#1'. Using the pattern for}%
\typeout{** the default language instead.}%
\else
\language=\csname l@#1\endcsname
\fi
#2}}
\providecommand{\BIBdecl}{\relax}
\BIBdecl

\bibitem{khaligh2010battery}
A.~Khaligh and Z.~Li, ``Battery, ultracapacitor, fuel cell, and hybrid energy
  storage systems for electric, hybrid electric, fuel cell, and plug-in hybrid
  electric vehicles: State of the art,'' \emph{IEEE Transactions on Vehicular
  Technology}, vol.~59, no.~6, pp. 2806--2814, 2010.

\bibitem{goodenough2013li}
J.~B. Goodenough and K.-S. Park, ``The li-ion rechargeable battery: a
  perspective,'' \emph{Journal of the American Chemical Society}, vol. 135,
  no.~4, pp. 1167--1176, 2013.

\bibitem{goodenough2015energy}
J.~B. Goodenough, ``Energy storage materials: a perspective,'' \emph{Energy
  Storage Materials}, vol.~1, pp. 158--161, 2015.

\bibitem{wang2017revisiting}
Y.~Wang, H.~Fang, L.~Zhou, and T.~Wada, ``Revisiting the state-of-charge
  estimation for lithium-ion batteries: A methodical investigation of the
  extended kalman filter approach,'' \emph{IEEE Control Systems Magazine},
  vol.~37, no.~4, pp. 73--96, 2017.

\bibitem{ramadass2004development}
P.~Ramadass, B.~Haran, P.~M. Gomadam, R.~White, and B.~N. Popov, ``Development
  of first principles capacity fade model for li-ion cells,'' \emph{Journal of
  the Electrochemical Society}, vol. 151, no.~2, p. A196, 2004.

\bibitem{wang2016lithium}
C.-Y. Wang, G.~Zhang, S.~Ge, T.~Xu, Y.~Ji, X.-G. Yang, and Y.~Leng,
  ``Lithium-ion battery structure that self-heats at low temperatures,''
  \emph{Nature}, vol. 529, no. 7587, pp. 515--518, 2016.

\bibitem{yang2018fast}
X.-G. Yang, G.~Zhang, S.~Ge, and C.-Y. Wang, ``Fast charging of lithium-ion
  batteries at all temperatures,'' \emph{Proceedings of the National Academy of
  Sciences}, vol. 115, no.~28, pp. 7266--7271, 2018.

\bibitem{yang2021challenges}
X.-G. Yang, T.~Liu, S.~Ge, E.~Rountree, and C.-Y. Wang, ``Challenges and key
  requirements of batteries for electric vertical takeoff and landing
  aircraft,'' \emph{Joule}, vol.~5, no.~7, pp. 1644--1659, 2021.

\bibitem{sun2020review}
P.~Sun, R.~Bisschop, H.~Niu, and X.~Huang, ``A review of battery fires in
  electric vehicles,'' \emph{Fire Technology}, vol.~56, pp. 1361--1410, 2020.

\bibitem{klein2012electrochemical}
R.~Klein, N.~A. Chaturvedi, J.~Christensen, J.~Ahmed, R.~Findeisen, and
  A.~Kojic, ``Electrochemical model based observer design for a lithium-ion
  battery,'' \emph{IEEE Transactions on Control Systems Technology}, vol.~21,
  no.~2, pp. 289--301, 2012.

\bibitem{weaver2020novel}
T.~Weaver, A.~Allam, and S.~Onori, ``A novel lithium-ion battery pack modeling
  framework-series-connected case study,'' in \emph{2020 American Control
  Conference (ACC)}.\hskip 1em plus 0.5em minus 0.4em\relax IEEE, 2020, pp.
  365--372.

\bibitem{tian2020nonlinear}
N.~Tian, H.~Fang, J.~Chen, and Y.~Wang, ``Nonlinear double-capacitor model for
  rechargeable batteries: Modeling, identification, and validation,''
  \emph{IEEE Transactions on Control Systems Technology}, vol.~29, no.~1, pp.
  370--384, 2020.

\bibitem{plett2015battery}
G.~L. Plett, \emph{Battery management systems, Volume I: Battery
  modeling}.\hskip 1em plus 0.5em minus 0.4em\relax Artech House, 2015, vol.~1.

\bibitem{he2011evaluation}
H.~He, R.~Xiong, and J.~Fan, ``Evaluation of lithium-ion battery equivalent
  circuit models for state of charge estimation by an experimental approach,''
  \emph{Energies}, vol.~4, no.~4, pp. 582--598, 2011.

\bibitem{johnson2002battery}
V.~Johnson, ``Battery performance models in advisor,'' \emph{Journal of Power
  Sources}, vol. 110, no.~2, pp. 321--329, 2002.

\bibitem{fang2016health}
H.~Fang, Y.~Wang, and J.~Chen, ``Health-aware and user-involved battery
  charging management for electric vehicles: Linear quadratic strategies,''
  \emph{IEEE Transactions on Control Systems Technology}, vol.~25, no.~3, pp.
  911--923, 2017.

\bibitem{lin2014lumped}
X.~Lin, H.~E. Perez, S.~Mohan, J.~B. Siegel, A.~G. Stefanopoulou, Y.~Ding, and
  M.~P. Castanier, ``A lumped-parameter electro-thermal model for cylindrical
  batteries,'' \emph{Journal of Power Sources}, vol. 257, pp. 1--11, 2014.

\bibitem{perez2017optimal}
H.~E. Perez, X.~Hu, S.~Dey, and S.~J. Moura, ``Optimal charging of li-ion
  batteries with coupled electro-thermal-aging dynamics,'' \emph{IEEE
  Transactions on Vehicular Technology}, vol.~66, no.~9, pp. 7761--7770, 2017.

\bibitem{biju2023battx}
N.~Biju and H.~Fang, ``Battx: An equivalent circuit model for lithium-ion
  batteries over broad current ranges,'' \emph{Applied Energy}, vol. 339, p.
  120905, 2023.

\bibitem{hussein2011review}
A.~A.-H. Hussein and I.~Batarseh, ``A review of charging algorithms for nickel
  and lithium battery chargers,'' \emph{IEEE Transactions on Vehicular
  Technology}, vol.~60, no.~3, pp. 830--838, 2011.

\bibitem{10155935}
Z.~Lu and S.~Mou, ``Variable sampling mpc via differentiable time-warping
  function,'' in \emph{2023 American Control Conference (ACC)}, 2023, pp.
  533--538.

\bibitem{klein2011optimal}
R.~Klein, N.~A. Chaturvedi, J.~Christensen, J.~Ahmed, R.~Findeisen, and
  A.~Kojic, ``Optimal charging strategies in lithium-ion battery,'' in
  \emph{Proceedings of the 2011 American Control Conference}.\hskip 1em plus
  0.5em minus 0.4em\relax IEEE, 2011, pp. 382--387.

\bibitem{xavier2015lithium}
M.~A. Xavier and M.~S. Trimboli, ``Lithium-ion battery cell-level control using
  constrained model predictive control and equivalent circuit models,''
  \emph{Journal of Power Sources}, vol. 285, pp. 374--384, 2015.

\bibitem{zou2017electrothermal}
C.~Zou, X.~Hu, Z.~Wei, and X.~Tang, ``Electrothermal dynamics-conscious
  lithium-ion battery cell-level charging management via state-monitored
  predictive control,'' \emph{Energy}, vol. 141, pp. 250--259, 2017.

\bibitem{zou2018model}
C.~Zou, C.~Manzie, and D.~Ne{\v{s}}i{\'c}, ``Model predictive control for
  lithium-ion battery optimal charging,'' \emph{IEEE/ASME Transactions on
  Mechatronics}, vol.~23, no.~2, pp. 947--957, 2018.

\bibitem{ouyang2018optimal}
Q.~Ouyang, J.~Chen, J.~Zheng, and H.~Fang, ``Optimal multiobjective charging
  for lithium-ion battery packs: A hierarchical control approach,'' \emph{IEEE
  Transactions on Industrial Informatics}, vol.~14, no.~9, pp. 4243--4253,
  2018.

\bibitem{fang2018optimal}
H.~Fang, C.~Depcik, and V.~Lvovich, ``Optimal pulse-modulated lithium-ion
  battery charging: Algorithms and simulation,'' \emph{Journal of Energy
  Storage}, vol.~15, pp. 359--367, 2018.

\bibitem{tian2020real}
N.~Tian, H.~Fang, and Y.~Wang, ``Real-time optimal lithium-ion battery charging
  based on explicit model predictive control,'' \emph{IEEE Transactions on
  Industrial Informatics}, vol.~17, no.~2, pp. 1318--1330, 2020.

\bibitem{azimi2022extending}
V.~Azimi, A.~Allam, and S.~Onori, ``Extending life of lithium-ion battery
  systems by embracing heterogeneities via an optimal control-based active
  balancing strategy,'' \emph{IEEE Transactions on Control Systems Technology},
  vol.~31, no.~3, pp. 1235--1249, 2023.

\bibitem{romagnoli2019feedback}
R.~Romagnoli, L.~D. Couto, A.~Goldar, M.~Kinnaert, and E.~Garone, ``A feedback
  charge strategy for li-ion battery cells based on reference governor,''
  \emph{Journal of Process Control}, vol.~83, pp. 164--176, 2019.

\bibitem{goldar2020low}
A.~Goldar, R.~Romagnoli, L.~D. Couto, M.~Nicotra, M.~Kinnaert, and E.~Garone,
  ``Low-complexity fast charging strategies based on explicit reference
  governors for li-ion battery cells,'' \emph{IEEE Transactions on Control
  Systems Technology}, vol.~29, no.~4, pp. 1597--1608, 2020.

\bibitem{couto2021faster}
L.~D. Couto, R.~Romagnoli, S.~Park, D.~Zhang, S.~J. Moura, M.~Kinnaert, and
  E.~Garone, ``Faster and healthier charging of lithium-ion batteries via
  constrained feedback control,'' \emph{IEEE Transactions on Control Systems
  Technology}, vol.~30, no.~5, pp. 1990--2001, 2021.

\bibitem{garone2017reference}
E.~Garone, S.~Di~Cairano, and I.~Kolmanovsky, ``Reference and command governors
  for systems with constraints: A survey on theory and applications,''
  \emph{Automatica}, vol.~75, pp. 306--328, 2017.

\bibitem{feng2024safe}
S.~Feng, R.~de~Castro, and I.~Ebrahimi, ``Safe battery control using
  cascade-control-barrier functions,'' \emph{IEEE Transactions on Control
  Systems Technology}, 2024.

\bibitem{galuppini2024efficient}
G.~Galuppini, M.~D. Berliner, H.~Lian, D.~Zhuang, M.~Z. Bazant, and R.~D.
  Braatz, ``Efficient computation of robust, safe, fast charging protocols for
  lithium-ion batteries,'' \emph{Control Engineering Practice}, vol. 145, p.
  105856, 2024.

\bibitem{dong2024optimal}
G.~Dong, Z.~Zhu, Y.~Lou, J.~Yu, L.~Wu, and J.~Wei, ``Optimal charging of
  lithium-ion battery using distributionally robust model predictive control
  with wasserstein metric,'' \emph{IEEE Transactions on Industrial
  Informatics}, vol.~20, no.~5, pp. 7630--7640, 2024.

\bibitem{li2023nonlinear}
Y.~Li, T.~Wik, Y.~Huang, and C.~Zou, ``Nonlinear model inversion-based output
  tracking control for battery fast charging,'' \emph{IEEE Transactions on
  Control Systems Technology}, vol.~32, no.~1, pp. 225--240, 2024.

\bibitem{zeng2021review}
Y.~Zeng, D.~Chalise, S.~D. Lubner, S.~Kaur, and R.~S. Prasher, ``A review of
  thermal physics and management inside lithium-ion batteries for high energy
  density and fast charging,'' \emph{Energy Storage Materials}, vol.~41, pp.
  264--288, 2021.

\bibitem{perez2017optimaljes}
H.~Perez, S.~Dey, X.~Hu, and S.~Moura, ``Optimal charging of li-ion batteries
  via a single particle model with electrolyte and thermal dynamics,''
  \emph{Journal of The Electrochemical Society}, vol. 164, no.~7, p. A1679,
  2017.

\bibitem{yin2021optimal}
Y.~Yin, Y.~Bi, Y.~Hu, and S.-Y. Choe, ``Optimal fast charging method for a
  large-format lithium-ion battery based on nonlinear model predictive control
  and reduced order electrochemical model,'' \emph{Journal of The
  Electrochemical Society}, vol. 167, no.~16, p. 160559, 2021.

\bibitem{storch2021temperature}
M.~Storch, J.~P. Fath, J.~Sieg, D.~Vrankovic, C.~Krupp, B.~Spier, and
  R.~Riedel, ``Temperature and lithium concentration gradient caused
  inhomogeneous plating in large-format lithium-ion cells,'' \emph{Journal of
  Energy Storage}, vol.~41, p. 102887, 2021.

\bibitem{goshtasbi2024evtol}
A.~Goshtasbi, S.~Han, R.~Zhao, and J.~Neubauer, ``Optimal charging with active
  thermal management for evtol aircraft battery packs,'' in \emph{2024 American
  Control Conference (ACC)}, 2024, pp. 695--700.

\bibitem{mohtat2021algorithmic}
P.~Mohtat, S.~Pannala, V.~Sulzer, J.~B. Siegel, and A.~G. Stefanopoulou, ``An
  algorithmic safety vest for li-ion batteries during fast charging,''
  \emph{IFAC-PapersOnLine}, vol.~54, no.~20, pp. 522--527, 2021.

\bibitem{hamednia2023optimal}
A.~Hamednia, N.~Murgovski, J.~Fredriksson, J.~Forsman, M.~Pourabdollah, and
  V.~Larsson, ``Optimal thermal management, charging, and eco-driving of
  battery electric vehicles,'' \emph{IEEE Transactions on Vehicular
  Technology}, vol.~72, no.~6, pp. 7265--7278, 2023.

\bibitem{tu2024system}
H.~Tu, X.~Lin, Y.~Wang, and H.~Fang, ``System identification for lithium-ion
  batteries with nonlinear coupled electro-thermal dynamics via bayesian
  optimization,'' \emph{arXiv preprint arXiv:2405.20219}, 2024.

\bibitem{andersson2019casadi}
J.~A. Andersson, J.~Gillis, G.~Horn, J.~B. Rawlings, and M.~Diehl, ``Casadi: a
  software framework for nonlinear optimization and optimal control,''
  \emph{Mathematical Programming Computation}, vol.~11, pp. 1--36, 2019.

\bibitem{wachter2006implementation}
A.~W{\"a}chter and L.~T. Biegler, ``On the implementation of an interior-point
  filter line-search algorithm for large-scale nonlinear programming,''
  \emph{Mathematical Programming}, vol. 106, pp. 25--57, 2006.

\bibitem{lowry2014concepts}
R.~Lowry, ``Concepts and applications of inferential statistics,'' 2014.

\bibitem{anderson2012optimal}
B.~D. Anderson and J.~B. Moore, \emph{Optimal filtering}.\hskip 1em plus 0.5em
  minus 0.4em\relax Courier Corporation, 2012.

\bibitem{wonham1968separation}
W.~M. Wonham, ``On the separation theorem of stochastic control,'' \emph{SIAM
  Journal on Control}, vol.~6, no.~2, pp. 312--326, 1968.

\bibitem{atassi1999separation}
A.~N. Atassi and H.~K. Khalil, ``A separation principle for the stabilization
  of a class of nonlinear systems,'' \emph{IEEE Transactions on Automatic
  Control}, vol.~44, no.~9, pp. 1672--1687, 1999.

\bibitem{georgiou2013separation}
T.~T. Georgiou and A.~Lindquist, ``The separation principle in stochastic
  control, redux,'' \emph{IEEE Transactions on Automatic Control}, vol.~58,
  no.~10, pp. 2481--2494, 2013.

\bibitem{park2020computationally}
S.~Park and C.~Ahn, ``Computationally efficient stochastic model predictive
  controller for battery thermal management of electric vehicle,'' \emph{IEEE
  Transactions on Vehicular Technology}, vol.~69, no.~8, pp. 8407--8419, 2020.

\bibitem{park2021model}
------, ``Model predictive control with stochastically approximated cost-to-go
  for battery cooling system of electric vehicles,'' \emph{IEEE Transactions on
  Vehicular Technology}, vol.~70, no.~5, pp. 4312--4323, 2021.

\bibitem{hajidavalloo2023nmpc}
M.~R. Hajidavalloo, J.~Chen, Q.~Hu, Z.~Song, X.~Yin, and Z.~Li, ``Nmpc-based
  integrated thermal management of battery and cabin for electric vehicles in
  cold weather conditions,'' \emph{IEEE Transactions on Intelligent Vehicles},
  vol.~8, no.~9, pp. 4208--4222, 2023.

\end{thebibliography}

\end{document}